\documentclass[letterpaper,11pt,spanish]{article}

 \title{La falacia del empate t\'ecnico electoral}
 \author{Arturo Erdely\thanks{Sitio personal en internet https://sites.google.com/site/arturoerdely}}
 \date{\small{Facultad de Estudios Superiores Acatl\'an \\
              Universidad Nacional Aut\'onoma de M\'exico \\
							\texttt{arturo.erdely@comunidad.unam.mx}\\}}

 \usepackage{amsthm}
 \usepackage{amssymb}
 \usepackage{amsmath}
 \usepackage{enumerate}
 \usepackage{graphicx}
 \usepackage{babel}

\newcommand{\prob}{\mathbb{P}}

\newcommand{\indic}{\textbf{\textsf{\large{1}}}}

\theoremstyle{plain}

\theoremstyle{definition}

\theoremstyle{remark}

\begin{document}
 
\maketitle

\begin{abstract}
  \noindent Se argumenta que el concepto de ``empate t\'ecnico'' en encuestas y conteos r\'apidos electorales no tiene fundamento probabil\'istico, y que en su lugar la incertidumbre asociada a dichos ejercicios estad\'isticos debiera expresarse en t\'erminos de una probabilidad de triunfo del candidato puntero.
\end{abstract}


\noindent \textbf{Palabras clave:} empate t\'ecnico, encuesta electoral, conteo r\'apido, probabilidad de triunfo.


\bigskip\bigskip

\noindent \underline{\'INDICE} \medskip

\noindent 1. Introducci\'on.\par
\noindent 2. ?`Qu\'e es ``empate t\'ecnico''?\par
\noindent 3. Contraejemplos.\par
\noindent 4. La elecci\'on presidencial de M\'exico en 2006.\par
\noindent 5. Conclusiones.

\bigskip

\section{Introducci\'on}

\noindent Un \textit{evento} es una aseveraci\'on (o proposici\'on l\'ogica) que resultar\'a verdadera o falsa una vez que se conozca el resultado del fen\'omeno (o experimento) al que se encuentra asociado. Si en un momento dado el resultado de dicho fen\'omeno es incierto, tambi\'en lo es la ocurrencia de (casi) cualquier evento asociado a \'el. De acuerdo a Lindley (2000) la disciplina conocida como \textit{Estad\'istica} tiene por objetivo central el estudio de la incertidumbre, particularmente la cuantificaci\'on y combinaci\'on de incertidumbres, auxili\'andose de la \textit{Teor\'ia de la Probabilidad} y con base en informaci\'on muestral:
\begin{quote}
  \textsl{``\ldots the statistician's role is to assist workers in other fields, the clients, who encounter uncertainty in their work. In practice, there is a restriction in that statistics is ordinarily associated with data; and it is the link between the uncertainty, or variability, in the data and that in the topic itself that has occupied statisticians [\ldots] A scientific approach would mean the \textbf{measurement of uncertainty}; for, to follow Kelvin, it is only by associating numbers with any scientific concept that the concept can be properly understood.''}
\end{quote} 

\medskip

\noindent Si $\,\mathcal{E}$ representa un evento de inter\'es cuya ocurrencia es incierta, desde un punto de vista estad\'istico se busca cuantificar probabil\'isticamente el grado de incertidumbre que al respecto se tiene con base en la informaci\'on disponible en un determinado momento en el tiempo. Lo usual es asociar a un evento de inter\'es $\,\mathcal{E}$ un n\'umero en el intervalo $\,[\,0\,,\,1\,]\,$ que se denota $\prob(\mathcal{E})$ y denomina \textit{probabilidad del evento $\,\mathcal{E}$}, bajo la convenci\'on de que valores m\'as cercanos a $1$ que a $0$ representan mayor cercan\'ia a la certidumbre de ocurrencia del evento, que valores m\'as cercanos a $0$ que a $1$ representan mayor cercan\'ia a la certidumbre de no ocurrencia del evento, y por tanto $\prob(\mathcal{E})=0{.}5$ representa el mayor nivel de incertidumbre. 

\medskip

\noindent Si consideramos, por ejemplo, la elecci\'on presidencial de M\'exico en el a\~no 2018 y es de inter\'es el evento $$\mathcal{E}\,=\, \text{``el candidato L\'opez Obrador gana la elecci\'on presidencial'',}$$ dependiendo de la informaci\'on que en cada momento del proceso electoral se tenga es posible la asignaci\'on de distintas probabilidades a la ocurrencia de dicho evento. Antes del d\'ia de la elecci\'on es posible recurrir a diversos tipos de \textit{encuestas electorales} consistentes en preguntar a un subconjunto (muestra) de los posibles votantes cu\'al es su preferencia electoral, realizar un an\'alisis estad\'istico de la informaci\'on recabada y hacer inferencias sobre el resultado de la elecci\'on. Dicho ejercicio involucra diversas fuentes de variabilidad que afectan la medici\'on de la incertidumbre de inter\'es: tama\~no y representatividad de la muestra de votantes, la honestidad en sus respuestas y la no respuesta, as\'i como los posibles cambios de opini\'on posteriores al levantamiento de la muestra, entre otros.

\medskip

\noindent El d\'ia de la elecci\'on tambi\'en es posible recurrir a otro tipo de ejercicios para inferir el resultado de la misma, antes de que varios d\'ias despu\'es se conozca el resultado del recuento total de votos emitidos: la \textit{encuesta de salida} y el \textit{conteo r\'apido.} La \textbf{encuesta de salida} consiste igualmente en preguntar a un subconjunto de las personas que ejercieron su voto por cu\'al candidato votaron justo al momento de salir de la casilla de votaci\'on, y en el an\'alisis estad\'istico de la informaci\'on recabada ser\'a nuevamente necesario considerar las fuentes de variabilidad ya mencionadas. El \textbf{conteo r\'apido} consiste en realizar una inferencia estad\'istica respecto al resultado de la elecci\'on con base en un subconjunto (muestra) del total de \textit{actas de escrutinio y c\'omputo}, que son documentos llenados y firmados por los funcionarios de cada casilla electoral en donde asentaron el resultado del conteo de votos en la casilla correspondiente. Gracias a esto, se eliminan algunas fuentes importantes de variabilidad que tienen las encuestas electorales, como es el caso de la deshonestidad, la no respuesta o el cambio de opini\'on, pero persiste la proveniente del tama\~no y \textit{representatividad de la muestra}, entendiendo por esto \'ultimo que la muestra analizada se ``parezca mucho'' en preferencias electorales al conjunto total de votaci\'on emitida, y esto es posible lograrlo con la adecuada aplicaci\'on de \textit{t\'ecnicas de muestreo} y una eficiente implementaci\'on log\'istica y tecnol\'ogica por parte de las autoridades electorales encargadas de dicho proceso.

\medskip

\noindent En tiempos de procesos electorales y ejercicios estad\'isticos para realizar inferencias sobre los posibles resultados, en M\'exico y algunos otros pa\'ises es com\'un que empresas encuestadoras, pol\'iticos, analistas y hasta autoridades electorales hablen de que un momento dado existe ``empate t\'ecnico'' entre dos candidatos punteros y que por tanto no es estad\'isticamente posible inferir un ganador. El presente art\'iculo tiene por objetivo argumentar que:
\begin{itemize}
  \item El empleo de la expresi\'on ``empate t\'ecnico'' no tiene sustento probabil\'istico y por tanto no debiera utilizarse en inferencias estad\'isticas derivadas de encuestas y conteos r\'apidos electorales.
	\item La incertidumbre sobre el posible resultado de una elecci\'on debiera expresarse mediante la estimaci\'on de la probabilidad de triunfo del candidato puntero.
\end{itemize}
Primero se analizar\'a lo que usualmente se entiende por ``empate t\'ecnico'' y se utilizar\'an argumentos intuitivos para criticar esta expresi\'on. Despu\'es se abordar\'an algunos contraejemplos te\'oricos para demostrar que la probabilidad de triunfo de un candidato puede ser significativamente elevada a pesar de existir ``empate t\'ecnico''. Posteriormente se analizar\'a el caso de la elecci\'on presidencial en M\'exico del a\~no 2006 por tratarse de un caso en que (aparentemente) el resultado del conteo r\'apido no permit\'ia una inferencia confiable.

\section{?`Qu\'e es ``empate t\'ecnico''?}

\noindent Se denotar\'a mediante la letra griega $\,\theta\,$ la proporci\'on de votos que obtendr\'a un candidato participante en una elecci\'on entre dos o m\'as candidatos, cantidad que en el lenguaje de la Estad\'istica se denomina \textit{par\'ametro} de inter\'es, cuyo valor es desconocido antes del recuento total de votos de dicha elecci\'on, y por tanto se desea recabar informaci\'on que sea estad\'isticamente analizable para realizar alg\'un tipo de \textit{inferencia estad\'istica} respecto a dicho par\'ametro.

\medskip

\noindent Una \textit{estimaci\'on puntual} de $\,\theta\,$ consiste en calcular un valor $\,\widehat{\theta}\,$ en el intervalo $\,[\,0\,,\,1\,]\,$ (por tratarse de una proporci\'on) con base en informaci\'on recabada para tal fin, de modo que pueda considerarse una buena aproximaci\'on (en alg\'un sentido estad\'istico) del valor desconocido $\,\theta.$ 

\medskip

\noindent Una \textit{estimaci\'on por intervalo} de $\,\theta\,$ consiste en calcular un intervalo $\,[\,a\,,\,b\,]\,$ de forma tal que el valor desconocido $\,\theta\,$ tenga una \textit{aceptablemente elevada} probabilidad de estar dentro de dicho intervalo. Respecto a esto \'ultimo es importante aclarar el uso de la expresi\'on ``aceptablemente elevada'': una estimaci\'on por intervalo para $\,\theta\,$ que no requiere esfuerzo alguno y que adem\'as tiene la probabilidad  m\'axima de $1$ (es decir, 100\%) de contener el verdadero valor de $\,\theta\,$ es justamente el intervalo $\,[\,0\,,\,1\,]\,,$ lo cual es tan cierto como in\'util. Es posible construir intervalos de estimaci\'on de modo que su longitud $\,b-a\,$ sea mucho menor, pero sacrificando un poco la probabilidad de que dicho intervalo contenga al verdadero valor de $\,\theta.$ ?`Cu\'anto sacrificar? Eso es totalmente arbitrario y su especificaci\'on ya no es un problema estad\'istico sino de criterio y decisi\'on de quien aplica la t\'ecnica. Por ejemplo, el \textit{Instituto Nacional Electoral} de M\'exico (INE, 2016) establece en su \textit{Reglamento de Elecciones}, art\'iculo 373, que para los conteos r\'apidos \textit{institucionales}\footnote{Por \textit{conteo r\'apido institucional} se entiende aqu\'el realizado por un \textit{Comit\'e T\'ecnico Asesor de los Conteos R\'apidos} de acuerdo con lo establecido en el art\'iculo 362 del Reglamento de Elecciones del INE (2016).} debe utilizarse un nivel de 95\%:
\begin{quote}
  \noindent\textsl{[\ldots] d) La muestra deber\'a dise\~narse con una \textbf{confianza}\footnote{En Estad\'istica, los t\'erminos \textit{confianza} y \textit{probabilidad} no son equivalentes en un sentido estricto, aunque en la pr\'actica se les interpreta como si lo fueran, situaci\'on que no es especialmente relevante discutir para los objetivos del presente art\'iculo.} de noventa y cinco por ciento, y con una precisi\'on tal, que genere certidumbre estad\'istica en el cumplimiento de los objetivos requeridos por el tipo de elecci\'on.}
\end{quote}

\noindent Respecto a la \textit{precisi\'on}, que denotaremos mediante una cantidad positiva $\,\varepsilon,$ \'esta se refiere a un \textit{margen de error} entre la cantidad estimada $\,\widehat{\theta}\,$ y la cantidad desconocida $\,\theta\,$ que se desea estimar, esto es, por ejemplo:
\begin{equation}\label{error}
   \prob(\,|\theta\,-\,\widehat{\theta}\,|\,\leq\,\varepsilon\,)\,=\,0{.}95\,,
\end{equation}
donde el valor a utilizar para $\,\varepsilon\,$ lo determina el usuario de la t\'ecnica estad\'istica. En el citado art\'iculo 373 del Reglamento de Elecciones del INE (2016), por ejemplo, solo se pide un valor para $\,\varepsilon\,$ ``que genere certidumbre estad\'istica en el cumplimiento de los objetivos requeridos por el tipo de elecci\'on'', dej\'andolo as\'i abierto al criterio y decisi\'on de quienes realicen el conteo r\'apido. Es importante destacar que el \textit{tama\~no de muestra} requerido para la encuesta o conteo r\'apido electoral, que denotaremos mediante un entero positivo $\,n,$ depende del valor $\,\varepsilon\,$ elegido, existiendo una \textit{relaci\'on inversa} entre ellos: a menor/mayor valor de $\,\varepsilon\,$ (esto es, a mayor/menor precisi\'on) se requiere un mayor/menor tama\~no de muestra $\,n.$ 

\medskip

\noindent La desigualdad en (\ref{error}) es equivalente a
\begin{equation}\label{error2}
  \widehat{\theta}\,-\,\varepsilon\,\leq\,\theta\,\leq\,\widehat{\theta}\,+\,\varepsilon\,,
\end{equation}
y nos permite la interpretaci\'on alternativa que, con probabilidad 95\%, la proporci\'on desconocida de votos $\,\theta\,$ estar\'a en el intervalo
\begin{equation}\label{intervalo}
  [\,a\,,\,b\,]\,\,=\,\,[\,\widehat{\theta}\,-\,\varepsilon\,,\,\widehat{\theta}\,+\,\varepsilon\,]
\end{equation}
donde $\,\widehat{\theta}\,$ es una estimaci\'on puntual de $\,\theta\,$ con base en una muestra de tama\~no $\,n\,$ y una precisi\'on (o margen de error) dado $\,\varepsilon.$ N\'otese que la longitud del intervalo resultante ser\'ia $\,b-a=2\varepsilon.$ As\'i, por ejemplo, si se desea $\,\varepsilon = 0{.}005\,$ (esto es un margen de error de 0.5\%) entonces la longitud del intervalo resultante ser\'ia del doble, esto es de $0{.}01$ (en porcentaje: 1\%).

\medskip

\noindent Considerando una contienda electoral en la que participan dos o m\'as candidatos, es posible realizar estimaciones por intervalo como en (\ref{intervalo}) de forma \textbf{individual} para cada candidato. Con toda intenci\'on se resalta la palabra \textbf{individual} ya que la t\'ecnica estad\'istica para la construcci\'on de dichos intervalos no considera la posible \textit{interacci\'on estad\'istica} \textbf{entre candidatos}. Si se desea entrar en comparaciones y realizar una inferencia estad\'istica respecto a qu\'e candidato ganar\'ia una elecci\'on, se debe recurrir a un procedimiento que incorpore adem\'as las posibles interacciones entre candidatos (ver discusi\'on en la siguiente secci\'on), y por tanto intervalos calculados de forma individual y que no reconocen dichas interacciones \textbf{no son intervalos estad\'isticamente comparables}.

\medskip

\noindent Desafortunadamente, lo mencionado en el p\'arrafo anterior es ignorado (con o sin intenci\'on) por parte de muchos de quienes realizan, analizan y difunden encuestas y conteos r\'apidos electorales, recurriendo a la siguiente mala \textit{praxis} estad\'istica: si $\,[\,a_2\,,\,b_2\,]\,$ y $\,[\,a_1\,,\,b_1\,]\,$ son dos estimaciones por intervalo de dos candidatos punteros y se cumple la relaci\'on $\,a_2\,<\,a_1\,<\,b_2\,<\,b_1\,$ (traslape de intervalos) entonces lo llaman ``\textbf{empate t\'ecnico}'', y en caso de que $\,b_2\,<\,a_1\,$ (no traslape de intervalos) dicen que hay una ``tendencia clara'' que favorece al candidato del intervalo $\,[\,a_1\,,\,b_1\,]\,.$\footnote{Aunque es muy poco probable, no ser\'ia imposible que ocurriera (por redondeo, por ejemplo) que $\,b_2=a_1\,$ en cuyo caso ser\'ia interesante saber qu\'e dir\'ian los partidarios de esta mala \textit{praxis} estad\'istica: ?`ser\'ia o no ``empate t\'ecnico''?}

\medskip

\noindent Aunque ser\'a en la siguiente secci\'on donde se argumente formalmente por qu\'e no tiene justificaci\'on probabil\'istica la expresi\'on ``empate t\'ecnico'', es posible adelantar algunas ideas intuitivas al respecto analizando posibles escenarios de traslape y no traslape de los intervalos de estimaci\'on para los dos candidatos punteros de una elecci\'on.

\medskip

\noindent Consid\'erese primero un escenario en el que existe un traslape de los intervalos estimados para los dos candidatos punteros, ver Cuadro 1. En tal caso por un traslape de tan solo una cent\'esima de punto porcentual (0.01\%) y existiendo una distancia entre las estimaciones puntuales de 0.69\% se declarar\'ia un ``empate t\'ecnico''.

\begin{table}[h]
\begin{center}
\begin{tabular}{|c|c|c|c|} \hline
  Candidato &  intervalo estimado (en \%)  & estimaci\'on puntual (en \%) & margen de error (en \%) \\ \hline
	   { }    &          { }                 &        { }           &      {  }        \\
	    A     & $[\,35{.}75\,,\,36{.}45\,]$  &        $36{.}10$     &     $0{.}35$     \\
	   { }    &          { }                 &        { }           &      {  }        \\ \hline
	   { }    &          { }                 &        { }           &      {  }        \\
			B     & $[\,35{.}06\,,\,35{.}76\,]$  &        $35{.}41$     &     $0{.}35$     \\
	   { }    &          { }                 &        { }           &      {  }        \\ \hline
\end{tabular}
\end{center}
\caption{Ejemplo de ``empate t\'ecnico'' por existir traslape de intervalos.}
\label{Escenario1}
\end{table}

\noindent Ahora consid\'erese un escenario en el que no existe un traslape de los intervalos estimados, ver Cuadro 2. En este caso, con una distancia entre las estimaciones puntuales de 0.71\% se declarar\'ia una ``tendencia favorable'' al candidato A, a pesar de que la diferencia entre las estimaciones puntuales es tan solo mayor en 2 cent\'esimas de punto porcentual respecto al caso de ``empate t\'ecnico'' cuando, como se analizar\'a en la siguiente secci\'on, la probabilidad de triunfo del candidato A sobre el candidato B es pr\'acticamente la misma con o sin un ``peque\~no'' traslape de intervalos.

\begin{table}[h]
\begin{center}
\begin{tabular}{|c|c|c|c|} \hline
  Candidato &  intervalo estimado (en \%)  & estimaci\'on puntual (en \%) & margen de error (en \%) \\ \hline
	   { }    &          { }                 &        { }           &      {  }        \\
	    A     & $[\,35{.}75\,,\,36{.}45\,]$  &        $36{.}10$     &     $0{.}35$     \\
	   { }    &          { }                 &        { }           &      {  }        \\ \hline
	   { }    &          { }                 &        { }           &      {  }        \\
			B     & $[\,35{.}04\,,\,35{.}74\,]$  &        $35{.}39$     &     $0{.}35$     \\
	   { }    &          { }                 &        { }           &      {  }        \\ \hline
\end{tabular}
\end{center}
\caption{Ejemplo de ``tendencia favorable'' al candidato A por no existir traslape de intervalos.}
\label{Escenario2}
\end{table}

\medskip

\noindent El uso injustificado de la expresi\'on ``empate t\'ecnico'' ya fue en alguna ocasi\'on se\~nalado, por ejemplo por Campos\footnote{Act. Roy Campos, presidente de \textit{Consulta Mitofsky}, empresa especializada en generar, analizar y presentar informaci\'on para el dise\~no de estrategias, estimaciones de proyecci\'on y evaluaci\'on de desempe\~no en M\'exico, Estados Unidos y Centroam\'erica. Fuente: www.consulta.mx [consultado el 13-Ene-2018]} y Penna (2004):
\begin{quote}
  \noindent\textsl{``Si nos apegamos a la teor\'ia estad\'istica, no existe una definici\'on --por lo menos al d\'ia en que se escribe este art\'iculo, octubre de 2004-- de lo que es un <<empate t\'ecnico>>. En M\'exico el t\'ermino <<empate t\'ecnico>> hizo su aparici\'on por primera vez en la televisi\'on nacional en una encuesta de salida llevada a cabo en febrero de 1999, la jornada electoral result\'o muy competida, pero un medio de comunicaci\'on a las 18:00 hrs., con un sentido m\'as period\'istico que t\'ecnico, se atrevi\'o a declarar ganador para posteriormente conforme transcurr\'ia la tarde modificar su anuncio y salir a declarar <<empate t\'ecnico>>. Finalmente el triunfo fue para un partido distinto al que se hab\'ia declarado.} \par\smallskip\noindent\textsl{[\ldots] El uso de <<empate t\'ecnico>> parece ocultar al culpable de no poder dar resultados, el m\'etodo. Los que lo usan parecen decir: <<la elecci\'on est\'a muy cerrada>> cuando eso no es cierto siempre. Algunos incluso simplemente revisan el dise\~no muestral, ven que en la metodolog\'ia dice <<5 por ciento de error>> y de ah\'i cualquier distancia menor entre ganador y segundo lugar lo consideran empate. Ese es un error. Imagine
simplemente una encuesta que diga que tiene 5 por ciento de error te\'orico y resulte en una distancia de 4 puntos entre primer y segundo lugar, alguien podr\'a pensar que es empate porque puede resultar en un ganador distinto, pero resulta que tambi\'en podría resultar en que la ventaja real fuera de 9 puntos, lo cual no refleja una contienda cerrada.''}
\end{quote}

\noindent El uso de la expresi\'on ``empate t\'ecnico'' persiste por lo menos hasta el a\~no en que se escribe el presente art\'iculo, como ejemplo basta citar, en el contexto de la elecci\'on presidencial de M\'exico en el a\~no 2018, la nota period\'istica de la Figura 1.

\begin{figure}[h]
\begin{center}
\includegraphics[width = 8cm, keepaspectratio]{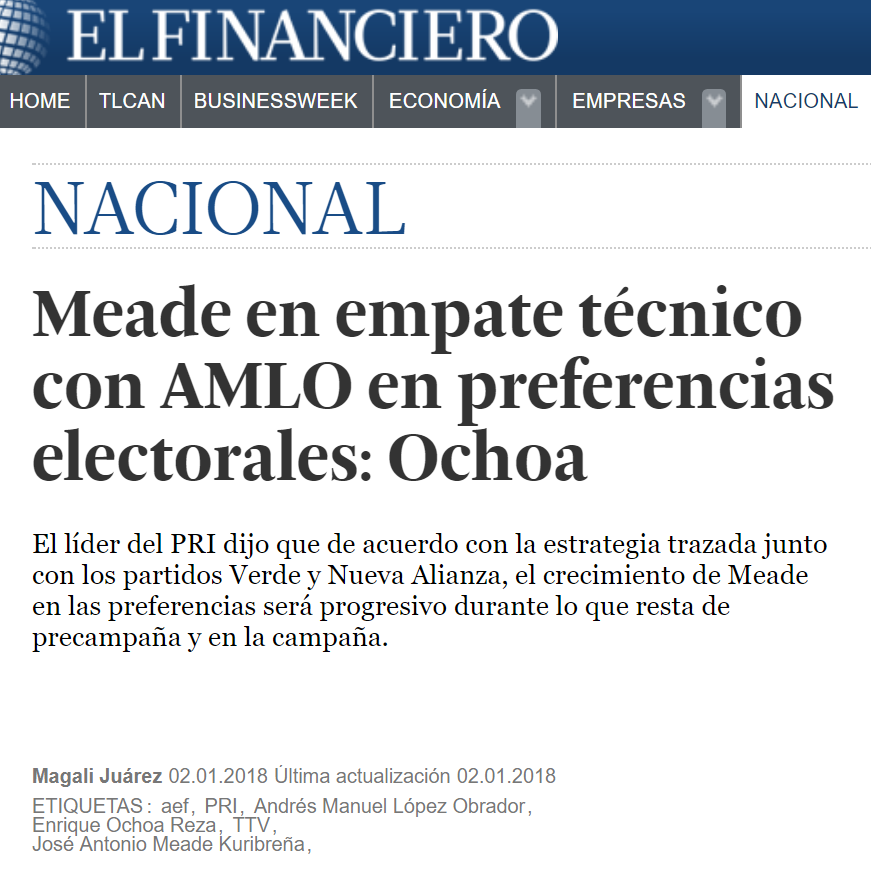}
\end{center}
\caption{Fuente: Peri\'odico \textit{El Financiero}, www.elfinanciero.com.mx consultado el 13-Ene-2018.}
\label{Ochoa}
\end{figure}

\noindent M\'as a\'un, hasta autoridades electorales llegan a utilizar la expresi\'on ``empate t\'ecnico'' y la difunden en informaci\'on dirigida a los electores, como es el caso del \textit{Instituto Electoral de Coahuila} respecto a la elecci\'on de gobernador en el a\~no 2017, que en su p\'agina de internet public\'o lo que se muestra en la Figura 2. En la siguiente secci\'on se aborda con rigor matem\'atico que el empleo de la expresi\'on ``empate t\'ecnico'' no tiene sustento probabil\'istico y por tanto no debiera utilizarse en inferencias estad\'isticas derivadas de encuestas y conteos r\'apidos electorales.

\begin{figure}[h]
\begin{center}
\includegraphics[width = 8cm, keepaspectratio]{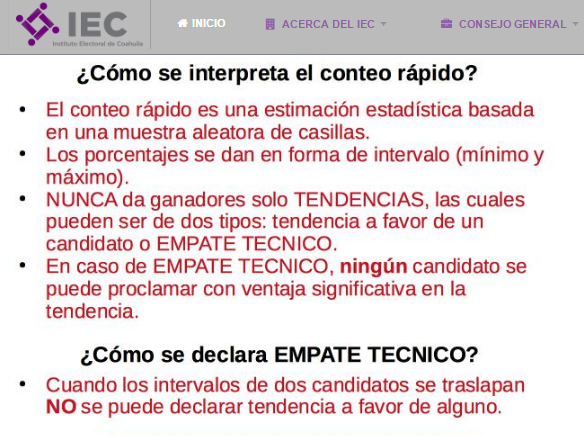}
\end{center}
\caption{Fuente: www.iec.org.mx/v1/index.php/conteo-rapido, consultado el 12-Jun-2017.}
\label{IECoahuila}
\end{figure}

\section{Contraejemplos}

\noindent Consid\'erese una elecci\'on en la que hay m\'as de dos candidatos contendientes, y que es de inter\'es calcular la probabilidad de triunfo entre los dos candidatos punteros. Las proporciones de votos de los dos candidatos punteros se representar\'an mediante las \textit{variables aleatorias} $\,X\,$ e $\,Y\,$ mientras que en una tercera variable aleatoria $Z$ se agregar\'a la proporci\'on restante de la votaci\'on emitida (otros candidatos registrados, votos anulados y votos por candidatos no registrados), de modo que debe cumplirse la condici\'on:
\begin{equation}\label{suma1}
  X\,+\,Y\,+\,Z\,\,=\,\,1\,,
\end{equation}
esto es, aunque antes del recuento total de votos las proporciones obtenidas mediante $X,Y$ y $Z$ son desconocidas, sea cual sea el resultado, su suma debe ser igual a 1 (es decir, equivalente al 100\% de la votaci\'on emitida). De (\ref{suma1}) se tiene que $\,Z=1-X-Y\,$ y por tanto una vez conocidos los valores para el vector aleatorio $\,(X,Y)\,$ el valor de $\,Z\,$ queda perfectamente determinado. Lo anterior implica que basta con tener un modelo probabil\'istico para el vector aleatorio $(X,Y)$ y con ello ser\'ia suficiente para realizar c\'alculos de probabilidades para $Z$ tambi\'en.

\medskip

\noindent Como la proporci\'on de votos a obtener por cada candidato es cualquier valor en el intervalo $\,[0,1]\,$ se considerar\'a a $\,(X,Y)\,$ como vector de variables aleatorias \textit{continuas} sobre dicho intervalo (lo cual implica lo mismo para $Z$). Consecuencia de (\ref{suma1}) se tiene tambi\'en que $\,0\,\leq\,X\,+\,Y\,\leq\,1\,$ y por tanto $\,0\leq\,X\,\leq\,1\,-\,Y,$ por lo que $\,\prob[\,(X,Y)\,\in\,\mathcal{S}\,]\,=\,1,$ donde $\mathcal{S}$ es el \textit{soporte} de $\,(X,Y)\,$ dado por el conjunto:
\begin{equation}\label{simplex}
  \mathcal{S}\,=\,\{(x,y)\,:\,0\leq y\leq 1\,,\,0\leq x\leq 1-y\}\,,
\end{equation}

\begin{figure}[h]
\begin{center}
\includegraphics[width = 8cm, keepaspectratio]{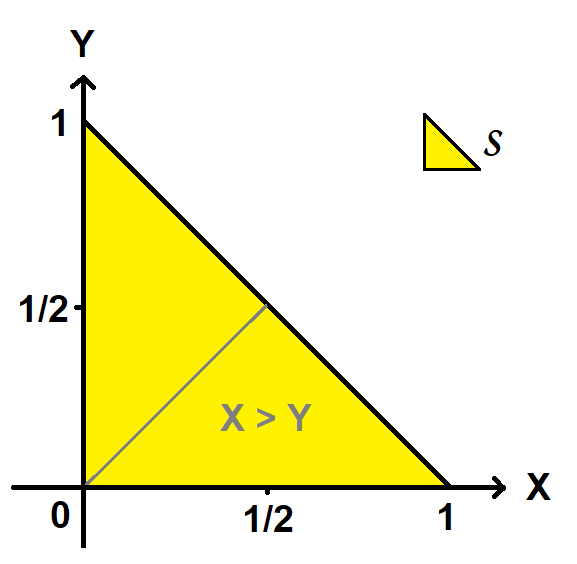}
\end{center}
\caption{Soporte $\mathcal{S}$ del vector aleatorio $(X,Y)$ y la regi\'on donde $X>Y.$}
\label{Simplex}
\end{figure}

\noindent como se ilustra en la Figura \ref{Simplex}. Como $(X,Y)$ representa las proporciones de votos de los dos candidatos punteros, es de inter\'es calcular la probabilidad de triunfo de alguno de ellos, por ejemplo $\,\prob(X>Y),$ misma que se calcula mediante:
\begin{equation}\label{ptriunfoX}
  \prob(X>Y)\,=\,\prob[\,(X,Y)\in\mathcal{S}_X\,]\,,\qquad\text{donde } \mathcal{S}_X=\{(x,y):0\leq y<1/2\,,\,y<x\leq 1-y\}\,\subset\,\mathcal{S}\,,
\end{equation}
ver Figura \ref{Simplex}. Toda la informaci\'on para hacer c\'alculo de probabilidades respecto al vector aleatorio $\,(X,Y)\,$ se encuentra en su funci\'on de distribuci\'on conjunta de probabilidades $\,F_{X,Y}(x,y)=\prob(X\leq x,Y\leq y),$ de donde se deducen adem\'as la funciones de distribuci\'on marginal (o individual) de las variables aleatorias $\,X\,$ e $\,Y,$ esto es $\,F_X(x)=\prob(X\leq x)=F_{X,Y}(x,+\infty)\,$ y $\,F_Y(y)=\prob(Y\leq y)=F_{X,Y}(+\infty,y).$ Como consecuencia del \textit{Teorema de Sklar} (1959) se tiene en este caso particular que existe una relaci\'on funcional \'unica entre la funci\'on de distribuci\'on conjunta $\,F_{X,Y}\,$ y las funciones de distribuci\'on marginales $\,F_X\,$ y $\,F_Y\,$ de la forma siguiente:
\begin{equation}\label{Sklar}
  F_{X,Y}(x,y)\,=\,\mathbf{C}\big(\,F_X(x)\,,\,F_Y(y)\,\big)\,,\qquad (x,y)\in\mathbb{R}^2,
\end{equation}
donde $\,\mathbf{C}:[0,1]^2\rightarrow[0,1]\,$ se denomina \textit{funci\'on c\'opula} asociada al vector aleatorio $\,(X,Y).$ Como las funciones de distribuci\'on marginales $\,F_X\,$ y $\,F_Y\,$ s\'olo explican el comportamiento probabil\'istico individual de las variables aleatorias $\,X\,$ e $\,Y\,$ \textbf{por separado}, entonces toda la informaci\'on acerca de la \textit{dependencia probabil\'istica} entre dichas variables aleatorias se encuentra justamente en $\,\mathbf{C}.$ As\'i, la funci\'on c\'opula que representa ausencia de dependencia (es decir, independencia) entre $\,X\,$ e $\,Y\,$ es $\,\mathbf{C}(u,v)=\mathbf{\Pi}(u,v)=uv,$ ya que es conocido resultado de la teor\'ia de la probabilidad que $\,X\,$ e $\,Y\,$ son independientes si y s\'olo si $\,\,F_{X,Y}(x,y)=F_X(x)F_Y(y)=\mathbf{\Pi}\big(F_X(x),F_Y(y)\big).$

\medskip

\noindent Tambi\'en, como consecuencia de las \textit{Cotas de Fr\'echet (1951) y Hoeffding(1940)} para funciones de distribuci\'on conjunta bivariadas, se tiene que:
\begin{equation}\label{cotasFH}
  \mathbf{W}\big(F_X(x),F_Y(y)\big)\,\leq\,F_{X,Y}(x,y)\,\leq\mathbf{M}\big(F_X(x),F_Y(y)\big)\,,
\end{equation}
con las funciones c\'opula $\,\mathbf{W}(u,v)=\max\{u+v-1,0\}\,$ y $\,\mathbf{M}(u,v)=\min\{u,v\}.$ Seg\'un se demuestra en Nelsen (2006), para un vector de variables aleatorias continuas $\,(X,Y)\,$ se tiene que su funci\'on c\'opula $\,\mathbf{C}=\mathbf{M}\,$ si y s\'olo si $\,Y\,$ es funci\'on estrictamente creciente de $\,X\,$ casi seguramente (en sentido probabil\'istico), y que $\,\mathbf{C}=\mathbf{W}\,$ si y s\'olo si $\,Y\,$ es funci\'on estrictamente decreciente de $\,X\,$ casi seguramente (en sentido probabil\'istico).

\medskip

\noindent Es inmediato verificar que para un vector de variables aleatorias continuas $\,(X,Y)\,$ con soporte en $\,\mathcal{S}\,$ (ver Figura \ref{Simplex}) \textbf{NO} es posible la independencia entre dichas variables ya que de serlo deber\'ia cumplirse para todo $\,0<x<1\,$ que $\,\prob(Y>x)=\prob(Y>x\,|\,X=x)\,$ y claramente $\,\prob(Y>x)>0\,$ mientras que la probabilidad $\,\prob(Y>x\,|\,X=x)=0\,$ como consecuencia del soporte $\,\mathcal{S}.$

\medskip

\noindent ?`Ser\'a posible que un vector aleatorio de variables aleatorias continuas $\,(X,Y)\,$ con soporte en $\,\mathcal{S}\,$ alcance (o se acerque mucho a) las cotas de Fr\'echet-Hoeffding, es decir que tenga por funci\'on c\'opula $\,\mathbf{C}=\mathbf{W}\,$ o bien $\,\mathbf{C}=\mathbf{M}$? La respuesta es en el sentido afirmativo: basta con que, por ejemplo, $\,X\,$ tenga una distribuci\'on de probabilidad continua uniforme sobre el intervalo $\,[0,1]\,$ y definir la variable aleatoria $\,Y\,=g_{\delta}(X)\,$ donde la funci\'on $\,g_{\delta}:[0,1]\rightarrow[0,1-\delta]\,$ con par\'ametro $\,0<\delta<1\,$ est\'a dada por:
\begin{eqnarray}
  g_{\delta}(x) &=& g_1(x)\indic_{\{0\,\leq\,x\,\leq\,\delta\}}\,+\,g_2(x)\indic_{\{\delta\,<\,x\,\leq\,1\}} \nonumber \\
	              &=& \frac{1-\delta}{\delta}\,x\indic_{\{0\,\leq\,x\,\leq\,\delta\}}\,+\,(1-x)\indic_{\{\delta\,<\,x\,\leq\,1\}}\,. \label{funciong}
\end{eqnarray}
Es relativamente sencillo demostrar\footnote{Se omiten los detalles por no ser especialmente relevantes para el objetivo del presente art\'iculo.} que la funci\'on de distribuci\'on de probabilidad marginal para $\,Y\,$ resulta ser en este caso continua uniforme tambi\'en pero sobre el intervalo $\,[0,1-\delta]\,$ y que la funci\'on c\'opula correspondiente a $\,(X,Y)\,$ en este caso resulta ser una familia param\'etrica $\,\{\mathbf{C}_{\delta}(u,v):0<\delta<1\}\,$ como en el ejemplo 3.3 en Nelsen (2006) donde $\,\lim_{\delta\rightarrow 0+}\mathbf{C}_{\delta}=\mathbf{W}\,$ y $\,\lim_{\delta\rightarrow 1-}\mathbf{C}_{\delta}=\mathbf{M}.$ Por lo tanto, se analizar\'an los casos extremos de dependencia para un vector de variables aleatorias continuas $\,(X,Y)\,$ con soporte en $\,\mathcal{S}.$

\bigskip

\noindent\textbf{\textit{Contraejemplo 1}}
\smallskip

\noindent Es posible tener $\,\mathbf{C}=\mathbf{W}\,$ si $\,Y\,$ es funci\'on estrictamente decreciente de $\,X.$ Definiendo $\,Y=1-X\,$ se cumple con ello y adem\'as dicha relaci\'on funcional es posible dentro de $\,\mathcal{S}\,$ (ver Figura \ref{Simplex}), que de hecho coincide con parte de la frontera de $\,\mathcal{S}\,$ que une los puntos $\,(0,1)\,$ y $\,(1,0).$ En este caso, la probabilidad de que $\,X\,$ triunfe sobre $\,Y\,$ estar\'ia dada por:
\begin{eqnarray}
  \prob(X>Y) &=& \prob(X>1-X)\,=\,\prob(X>1/2)\,=\,1-F_X(1/2)\,, \label{Wtriunfo1} \\
	           &=& \prob(1-Y>Y)\,=\,\prob(Y<1/2)\,=\,F_Y(1/2)\,, \label{Wtriunfo2}
\end{eqnarray}
donde 
\begin{equation}\label{FYW}
  F_Y(y)\,=\,\prob(Y\leq y)\,=\,\prob(1-X\leq y)\,=\,\prob(X\geq 1-y)\,=\,1-F_X(1-y)\,.
\end{equation}

\noindent De (\ref{Wtriunfo1}) se desprende que, para cualquier valor $\,0<\gamma<1,$ la probabilidad $\,\prob(X>Y)=1-\gamma\,$ si y s\'olo si $\,F_X(1/2)=\gamma,$ y de (\ref{Wtriunfo2}) se desprende que $\,\prob(X>Y)=1-\gamma\,$ si y s\'olo si $\,F_Y(1/2)=1-\gamma.$ Esto implicar\'ia que $\,[\,1/2\,,\,1\,]\,$ sea un intervalo de probabilidad $100(1-\gamma)\%$ para $\,X\,$ y que $\,[\,0\,,\,1/2\,]\,$ sea un intervalo de probabilidad $100(1-\gamma)\%$ para $\,Y,$ esto es:
\begin{equation}\label{FXFY}
  F_X(1)\,-\,F_X(1/2)\,=\,1-\gamma\,=\,F_Y(1/2)\,-\,F_Y(0)\,.
\end{equation}
Como $\,F_X(1/2)=\gamma,$ si $F_X$ es estrictamente creciente entonces por continuidad de $\,F_X\,$ existe un valor $\,0<x_{\gamma/2}<\frac{1}{2}\,$ tal que $\,F_X(x_{\gamma/2})=\gamma/2\,$ y an\'alogamente existe un valor $\,\frac{1}{2}<x_{1-\gamma/2}<1\,$ tal que $\,F_X(x_{1-\gamma/2})=1-\gamma/2\,,$ y por tanto:
\begin{equation}\label{FXFY2}
  F_X(x_{1-\gamma/2})-F_X(x_{\gamma/2})\,=\,(1-\gamma/2)-\gamma/2\,=\,1-\gamma\,,\qquad\text{con } 0<x_{\gamma/2}<1/2<x_{1-\gamma/2}<1\,.
\end{equation}
Por argumentos an\'alogos, tambi\'en es posible:
\begin{equation}\label{FXFY3}
  F_Y(y_{1-\gamma/2})-F_Y(y_{\gamma/2})\,=\,(1-\gamma/2)-\gamma/2\,=\,1-\gamma\,,\qquad\text{con } 0<y_{\gamma/2}<1/2<y_{1-\gamma/2}<x_{1-\gamma/2}\,,
\end{equation}
por lo que de (\ref{FXFY2}) y (\ref{FXFY3}) se concluye que es posible construir para $\,X\,$ un intervalo $\,[\,x_{\gamma/2}\,,\,x_{1-\gamma/2}\,]\,$ de probabilidad $100(1-\gamma)\%$ y para $\,Y\,$ un intervalo $\,[\,y_{\gamma/2}\,,\,y_{1-\gamma/2}\,]\,$ tambi\'en de probabilidad $100(1-\gamma)\%$ que tengan intersecci\'on no vac\'ia $\,[\,x_{\gamma/2}\,,\,y_{1-\gamma/2}]\,$ y con probabilidad de triunfo para $\,X\,$ sobre $\,Y\,$ igual a $\,\prob(X>Y)=1-\gamma.$ As\'i, por ejemplo, es posible tomar un valor tan peque\~no como $\gamma=0{.}0001$ y con ello $\,X\,$ tendr\'ia una probabilidad de triunfo sobre $\,Y\,$ de $99{.}99\%$ a pesar de existir intersecci\'on de los intervalos individualmente calculados. $\quad_{\square}$

\bigskip

\noindent\textbf{\textit{Contraejemplo 2}}
\smallskip

\begin{figure}[h]
\begin{center}
\includegraphics[width = 8cm, keepaspectratio]{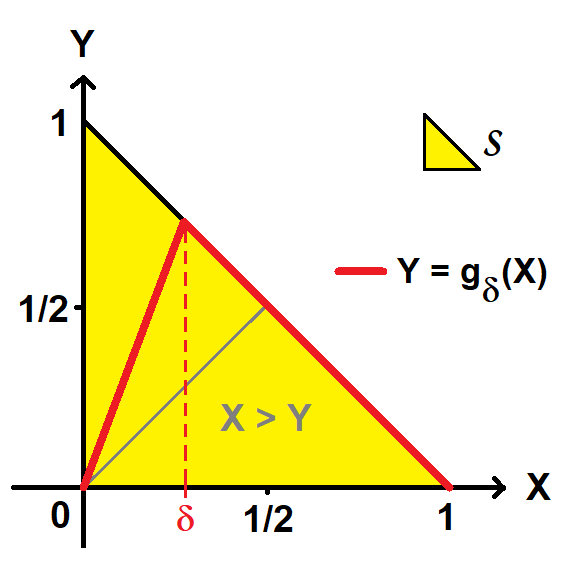}
\end{center}
\caption{En color rojo gr\'afica de $Y=g_{\delta}(X).$}
\label{Simplex2}
\end{figure}

\noindent Para que la funci\'on c\'opula $\,\mathbf{C}\,$ correspondiente a $\,(X,Y)\,$ fuese igual a $\,\mathbf{M}\,$ se requerir\'ia que $Y=\varphi(X)$ casi seguramente, donde $\varphi$ es una funci\'on estrictamente creciente, lo cual no es posible sobre $\,\mathcal{S}\,$ ya que necesariamente existir\'ia un valor $\,0<x_0<1\,$ tal que $\,\varphi(x_0)>1-x_0,$ es decir $\,Y\,$ reportar\'ia valores inadmisibles (fuera de $\mathcal{S}$). Pero es posible lograr que $\,\mathbf{C}\,$ se aproxime tanto como se desee a $\,\mathbf{M}\,$ mediante $\,Y=g_{\delta}(X)\,$ con $\,g_{\delta}\,$ definida como en (\ref{funciong}), esto es:
\begin{eqnarray}
  Y = g_{\delta}(X) &=& g_1(X)\indic_{\{0\,\leq\,X\,\leq\,\delta\}}\,+\,g_2(X)\indic_{\{\delta\,<\,X\,\leq\,1\}} \nonumber \\
	                  &=& \frac{1-\delta}{\delta}\,X\indic_{\{0\,\leq\,X\,\leq\,\delta\}}\,+\,(1-X)\indic_{\{\delta\,<\,X\,\leq\,1\}} \label{funciongYX}
\end{eqnarray}
con par\'ametro $\,0<\delta<1.$ El soporte de $\,Y\,$ ser\'ia el intervalo $\,[\,0,1-\delta\,]\,$ y su funci\'on de distribuci\'on para $\,0\leq y\leq1-\delta\,$ ser\'ia:
\begin{eqnarray}
  F_Y(y) &=& \prob(Y\leq y)\,=\,\prob(X\leq g_1^{-1}(y))\,+\,\prob(X\geq g_2^{-1}(y)) \nonumber \\
	       &=& F_X(\delta y/(1-\delta))\,+\,1\,-\,F_X(1-y)\,. \label{funcionYX2}
\end{eqnarray}
Con par\'ametro $\,\delta\geq 1/2\,$ la poligonal $\,Y=g_{\delta}(X)\,$ quedar\'ia totalmente contenida en $\,S_X\,,$ ver (\ref{ptriunfoX}), y en tal caso $\,\prob(X>Y)=1.$ Si se desea $\,\prob(X>Y)=1-\gamma\,$ para alg\'un valor dado $\,0<\gamma<1\,$ ser\'ia entonces necesario utilizar alg\'un valor $\,0<\delta<1/2\,,$ ver Figura \ref{Simplex2}, en cuyo caso se tendr\'ia que $\,1-\gamma=\prob(X>Y)=\prob(X>1/2)=1-F_X(1/2)\,$ y por lo tanto $\,F_X(1/2)=\gamma.$

\noindent Por otro lado, aplicando (\ref{funcionYX2}) se obtiene que:
\begin{equation}\label{funcionYX3}
  F_Y(1/2)=F_X(\delta/2(1-\delta))+1-F_X(1-1/2)=F_X(\delta/2(1-\delta))+1-\gamma=1-\gamma^*>1-\gamma\,,
\end{equation}
y aplicando argumentos an\'alogos a los del Contraejemplo 1 nuevamente es posible construir para $\,X\,$ un intervalo $\,[\,x_{\gamma/2}\,,\,x_{1-\gamma/2}\,]\,$ de probabilidad $100(1-\gamma)\%$ y para $\,Y\,$ un intervalo $\,[\,y_{\gamma^*/2}\,,\,y_{1-\gamma^*/2}\,]\,$ de probabilidad $100(1-\gamma^*)\%>100(1-\gamma)\%$ que tengan intersecci\'on no vac\'ia $\,[\,x_{\gamma/2}\,,\,y_{1-\gamma^*/2}]\,$ y con probabilidad de triunfo para $\,X\,$ sobre $\,Y\,$ igual a $\,\prob(X>Y)=1-\gamma.$ As\'i, por ejemplo, es posible tomar un valor tan peque\~no como $\gamma=0{.}0001$ y se tendr\'ia nuevamente que $\,X\,$ tendr\'ia una probabilidad de triunfo sobre $\,Y\,$ de $99{.}99\%$ a pesar de existir intersecci\'on de los intervalos individualmente calculados. $\quad_{\square}$

\bigskip

\noindent\textbf{\textit{Contraejemplo 3}}
\smallskip

\noindent En los dos contraejemplos anteriores se analizaron los casos extremos de dependencia que corresponden a las cotas inferior y superior de Fr\'echet-Hoeffding. Ahora se analizar\'an casos intermedios de dependencia para un vector de variables aleatorias continuas $\,(X,Y)\,$ con soporte contenido en $\,\mathcal{S}.$ Consid\'erese el caso de una elecci\'on en la que ninguno de los dos candidatos punteros alcanzar\'ia mayor\'ia absoluta de votos, esto es que $\,\prob\big((X,Y)\in[\,0\,,\,1/2\,]^2\big)=1.$ En este caso la probabilidad de triunfo de $\,X\,$ sobre $\,Y\,$ estar\'ia dada por:
\begin{equation}\label{C3A}
  \prob(X>Y)\,=\,\prob\big((X,Y)\in\mathcal{S}_X\cap[\,0\,,\,1/2\,]^2\big)\,=\,\prob(0\leq X\leq 1/2\,,\,0\leq Y<X)\,.
\end{equation}

\noindent Si $B$ es una variable aleatoria continua con distribuci\'on de probabilidad \textit{Beta} con par\'ametros $\,\alpha>0\,$ y $\,\beta>0,$ definiendo $\,X=B/2\,$ se logra que $\,\prob(0\leq X\leq 1/2)=1,$ y la relaci\'on entre la funci\'on de distribuci\'on de $\,X\,$ y la de $\,B\,$ estar\'ia dada por
\begin{equation}\label{C3B}
  F_X(x\,|\,\alpha,\beta)\,=\,\prob(X\leq x)\,=\,\prob(B\leq 2x)\,=\,F_B(2x\,|\,\alpha,\beta)\,.
\end{equation}

\noindent De forma an\'aloga a los dos contraejemplos anteriores, dados valores $\,0<\gamma<1\,$ y $\,0<2\varepsilon<x_{1-\gamma/2}<1/2\,$ bastar\'ia encontrar valores $\,\alpha>1\,$ y $\,\beta>1\,$\footnote{Las restricciones $\alpha>1$ y $\beta>1$ garantizan que la funci\'on de densidad de probabilidades \textit{Beta} tenga un m\'aximo y forma ``acampanada''.} tales que:
\begin{equation}\label{C3C}
  F_X(x_{1-\gamma/2}\,|\,\alpha,\beta)=1-\gamma/2 \qquad\text{y}\qquad F_X(x_{\gamma/2}\,|\,\alpha,\beta)=\gamma/2\,,
\end{equation}
donde $x_{\gamma/2}=x_{1-\gamma/2}-2\varepsilon\,.$ Lo anterior nos arroja un intervalo $\,[\,x_{\gamma/2}\,,\,x_{1-\gamma/2}\,]\,$ de longitud $\,2\varepsilon\,$ (esto es con margen de error $\varepsilon$) y de probabilidad $100(1-\gamma)\%$ para $\,X.$ A partir de este intervalo se puede construir uno similar para $\,Y\,$ que tenga un traslape deseado $\,0<\tau<2\varepsilon\,:$ t\'omese $\,y_{1-\gamma/2}=x_{\gamma/2}+\tau\,$ y apl\'iquese el mismo procedimiento (\ref{C3C}), con lo que se obtendr\'a un intervalo $\,[\,y_{\gamma/2}\,,\,y_{1-\gamma/2}\,]\,$ de longitud $\,2\varepsilon\,$ (esto es con margen de error $\varepsilon$) y de probabilidad $100(1-\gamma)\%$ para $\,Y\,$ que adem\'as tiene un traslape de longitud $\,\tau\,$ con el intervalo obtenido para $\,X.$

\medskip

\noindent Los dos intervalos con traslape para $\,X\,$ e $\,Y\,$ se obtienen \'unicamente a partir de las distribuciones de probabilidad marginales de cada variable aleatoria, sin considerar a\'un la posible dependencia entre ellas, misma que se proceder\'a a agregar ahora. Si $\,F_X(x\,|\,\alpha_1,\beta_1)\,$ y $\,F_Y(y\,|\,\alpha_2,\beta_2)\,$ son las funciones de distribuci\'on marginales obtenidas para lograr los intervalos traslapados anteriores, entonces por Teorema de Sklar (1959) la funci\'on de distribuci\'on conjunta de $\,(X,Y)\,$ es de la forma:
\begin{equation}\label{C3D}
  F_{X,Y}(x,y)\,=\,\mathbf{C}\big(\,F_X(x\,|\,\alpha_1,\beta_1)\,,\,F_Y(y\,|\,\alpha_2,\beta_2)\,\big)
\end{equation}
donde $\,\mathbf{C}\,$ es la funci\'on c\'opula que contiene toda la informaci\'on sobre la dependencia entre las variables aleatorias $\,X\,$ e $\,Y.$ Para el prop\'osito del presente contraejemplo se utilizar\'a (entre muchas opciones disponibles) una funci\'on c\'opula  perteneciente a la familia Frank (1979) por ser \'esta una familia param\'etrica $\,\{\mathbf{C}_{\delta}:\delta\in\,]-\infty,+\infty\,[\,\setminus\{0\}\}\,$ que abarca desde dependencias que se aproximan tanto como se desee a las cotas extremas de dependencia de Fr\'echet-Hoeffding ($\lim_{\delta\rightarrow -\infty}\mathbf{C}_{\delta}=\mathbf{W}\,$ y $\,\lim_{\delta\rightarrow +\infty}\mathbf{C}_{\delta}=\mathbf{M}$) hasta dependencias que pueden aproximarse a la independencia tanto como se desee ($\lim_{\delta\rightarrow 0}\mathbf{C}_{\delta}=\mathbf{\Pi}$). Bajo esta familia de c\'opulas, a mayor valor positivo del par\'ametro $\,\delta\,$ se obtiene una mayor dependencia positiva, a menor valor negativo se obtiene mayor dependencia negativa, y conforme el par\'ametro se acerque a cero se acercar\'a m\'as a la independencia. Una vez especificado (\ref{C3D}) se calcula la probabilidad de triunfo de $\,X\,$ sobre $\,Y\,$ mediante (\ref{C3A}).

\medskip

\noindent La soluci\'on del sistema de ecuaciones (\ref{C3C}) tiene que aproximarse num\'ericamente, y el c\'alculo de la probabilidad de triunfo (\ref{C3A}) se aproxima v\'ia simulaci\'on. En este caso se hizo por medio del lenguaje de programaci\'on \texttt{R} (2017) y el paquete \texttt{copula} de Hofert \textit{et al.}(2017). El c\'odigo de programaci\'on se anexa en el Ap\'endice, as\'i como un enlace para descarga del mismo, para fines de reproducibilidad de los c\'alculos. 

\medskip

\begin{figure}[t]
\begin{center}
\includegraphics[width = 8cm, keepaspectratio]{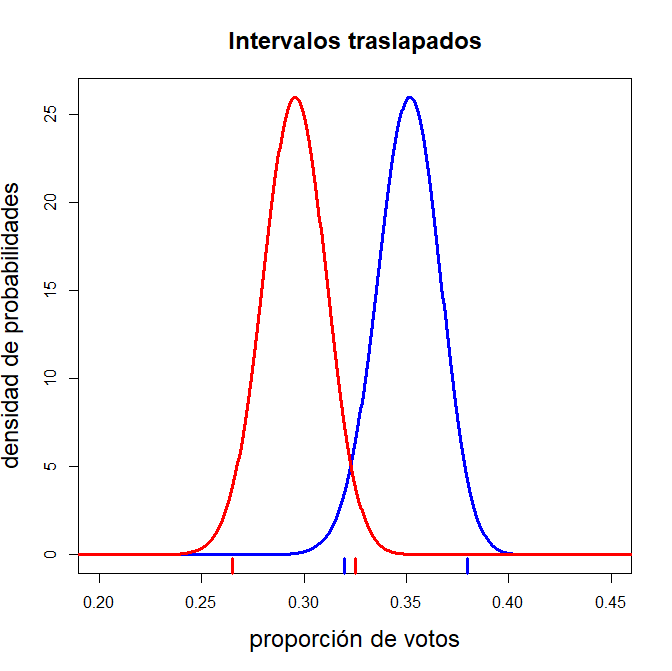}
\end{center}
\caption{Funciones de densidad de probabilidad marginal para $X$ (azul) e $Y$ (rojo) junto con intervalos de probabilidad $95\%$, margen de error de $3\%$ y traslape de $0{.}5\%.$}
\label{intervalos}
\end{figure}

\noindent Se utilizaron valores $\,\gamma=0{.}05,$ $x_{1-\gamma/2}=0{.}38,$ $\tau=0{.}005\,$ y $\,\varepsilon=0{.}03.$ Lo anterior genera intervalos de probabilidad $95\%$ con margen de error de $3\%$ y traslape de intervalos de $0{.}5\%$ como sigue:
\begin{equation}\label{C3E}
  Y\,:\,[\,26{.}5\%\,,\,32{.}5\%\,] \qquad\qquad X\,:\,[\,32{.}0\%\,,\,38{.}0\%\,]\,,
\end{equation}
con funciones de densidad de probabilidad marginal como se ilustra en la Figura \ref{intervalos}.

\medskip

\begin{figure}[t]
\begin{center}
\includegraphics[width = 8cm, keepaspectratio]{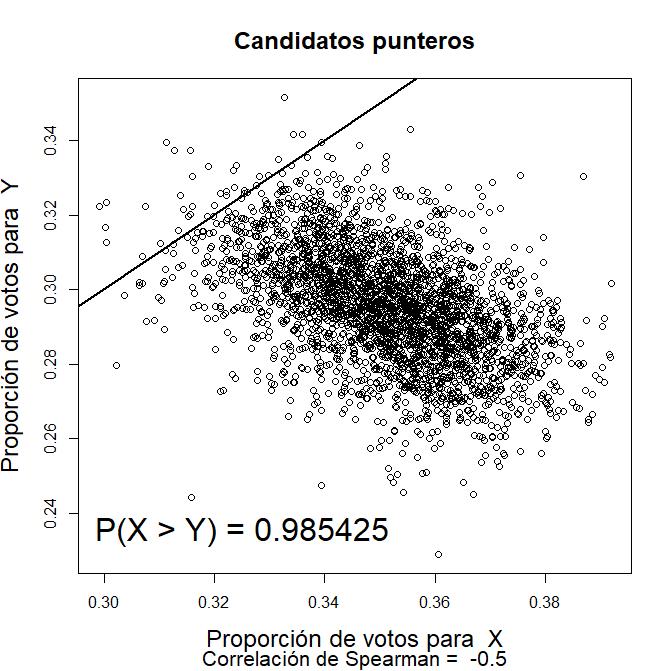}
\end{center}
\caption{Parte de las simulaciones de $(X,Y)$ con c\'opula Frank y correlaci\'on de Spearman igual a $-0{.}5.$ La parte inferior a la l\'inea recta corresponde a escenarios donde $X>Y.$}
\label{ejemploTriunfo}
\end{figure}

\noindent A las funciones de densidad de probabilidad marginal obtenidas, se agreg\'o la funci\'on c\'opula Frank con distintos valores de \textit{correlaci\'on de Spearman}\footnote{La correlaci\'on de Spearman es un valor en el intervalo $[-1,+1]$ cuya interpretaci\'on es muy similar a la correlaci\'on lineal o de Pearson, con la ventaja de que solo depende de la c\'opula subyacente, no se contamina por el comportamiento de las distribuciones marginales y no depende de la existencia de momentos. Ver Embrechts \textit{et al.}(1999).} y se procedi\'o a estimar v\'ia simulaci\'on la probabilidad de triunfo de $\,X\,$ sobre $\,Y,$ ver Cuadro \ref{probsTriunfo}. El resultado de la simulaci\'on para el caso particular donde la correlaci\'on de Spearman es igual a $-0{.}5$ se ilustra en la Figura \ref{ejemploTriunfo}.

\bigskip
\begin{table}[h]
\begin{center}
\begin{tabular}{|c|c|} \hline
  Correlaci\'on de Spearman & Probabilidad de triunfo de $X$ (en \%) \\ \hline
	      -0{.}9              &     97{.}13  \\ \hline
				-0{.}5              &     98{.}54  \\ \hline
				-0{.}2              &     99{.}14  \\ \hline
				 0{.}0              &     99{.}44  \\ \hline
				+0{.}3              &     99{.}77  \\ \hline
				+0{.}6              &     99{.}95  \\ \hline
				+0{.}9              &     99{.}99  \\ \hline
\end{tabular}
\end{center}
\caption{Probabilidad de triunfo de $X$ sobre $Y$ bajo distintos valores de correlaci\'on de Spearman, con intervalos de probabilidad $95\%,$ traslapados en $0{.}5\%$ y margen de error de $3\%$.}
\label{probsTriunfo}
\end{table}

\medskip

\noindent Como puede apreciarse en el Cuadro \ref{probsTriunfo}, a pesar de existir un traslape en los intervalos individuales para los candidatos punteros, bajo distintos niveles de dependencia (negativos, independencia, positivos) la probabilidad de triunfo de un candidato sobre el otro es siempre superior al nivel 95\% con el que fueron construidos los intervalos.

\medskip

\noindent Se ver\'a ahora que la probabilidad de triunfo puede resultar casi la misma con un m\'inimo traslape de intervalos que sin \'el. Basta utilizar los valores $\,\tau_1=+0{.}0001$ (traslape) y $\,\tau_2=-0{.}0001$ (no traslape) con los dem\'as valores igual que en lo anterior. Los resultados se resumen en el Cuadro \ref{tantito} y deja claro que un m\'inimo traslape de intervalos versus una m\'inima separaci\'on no cambia significativamente la probabilidad de triunfo de un candidato sobre otro. $\qquad_{\square}$

\bigskip
\begin{table}[h]
\begin{center}
\begin{tabular}{|c|c|c|c|} \hline
         Candidato $Y$        &        Candidato $X$        &  ?`Traslape?  &  Probabilidad de que triunfe $X$ (en \%)  \\ \hline
	$[\,26{.}01\,,\,32{.}01\,]$ & $[\,32{.}00\,,\,38{.}00\,]$ & \textbf{S\'i} &      99{.}19 \\ \hline
	$[\,25{.}99\,,\,31{.}99\,]$ & $[\,32{.}00\,,\,38{.}00\,]$ &  \textbf{No}  &      99{.}21 \\ \hline
\end{tabular}
\end{center}
\caption{Probabilidad de triunfo de $X$ sobre $Y$ con correlaci\'on de Spearman $-0{.}5$, con intervalos de probabilidad $95\%,$ margen de error de $3\%$, traslapados en un caso y en otro no.}
\label{tantito}
\end{table}

\bigskip

\noindent En resumen, los tres contraejemplos anteriores demuestran que es posible tener elevadas probabilidades de triunfo de un candidato puntero a pesar de que exista alg\'un traslape de los intervalos calculados por separado para los dos candidatos punteros, y que inclusive la probabilidad de triunfo puede resultar esencialmente la misma con un m\'inimo traslape de intervalos que sin \'el.

\section{La elecci\'on presidencial de M\'exico en 2006}

\noindent Se analiza este caso por tratarse de un ejemplo en el que, derivado de un conteo r\'apido a cargo de una autoridad electoral \'esta decide, con asesor\'ia de especialistas en estad\'istica, no emitir una conclusi\'on al respecto la noche del d\'ia de la elecci\'on. De acuerdo con Eslava (2006), quien form\'o parte del \textbf{C}omit\'e \textbf{T}\'ecnico \textbf{A}sesor para la \textbf{R}ealizaci\'on de \textbf{C}onteos \textbf{R}\'apidos (CTARCR) en la elecci\'on presidencial de M\'exico el 2 de julio del a\~no 2006:
\begin{quote}
  \noindent\textsl{``A pesar de que se obtuvieron estimaciones puntuales de los porcentajes, dos de los intervalos de confiabilidad asociados no se separaron por m\'as de 0.006, margen acordado para poder proporcionar los resultados la misma noche. Por esta raz\'on no se difundieron las cifras estimadas, sino solamente el aviso de que estad\'isticamente no era posible distinguir con un alto grado de confiabilidad a un candidato ganador [\ldots] El conteo r\'apido, por ser de car\'acter institucional y cuyos resultados ser\'ian la base de un comunicado difundido a la ciudadan\'ia la misma noche de la elecci\'on, estuvo sujeto a acuerdos previos y a restricciones como guardar la confidencialidad de la muestra; realizar estimaciones por intervalos de cuando menos una confiabilidad de 95\%; que para identificar un partido o coalici\'on ganadora, los intervalos asociados a los porcentajes estimados de votos para los partidos mayoritarios deber\'an distar en al menos 0.6\% (0.006)\ldots''}
\end{quote}

\noindent En el \textit{Informe sobre las actividades del Comit\'e T\'ecnico Asesor para la Realizaci\'on de Conteos R\'apidos} (CTARCR) del entonces \textit{Instituto Federal Electoral} (IFE, 2006, 147 p\'aginas) se incluyen tres metodolog\'ias estad\'isticas para la estimaci\'on por intervalos (denominados m\'etodos \textit{Robusto}, \textit{Cl\'asico} y \textit{Bayesiano}) as\'i como la t\'ecnica de muestreo utilizada, entre otras cuestiones. Sin embargo, no aparece metodolog\'ia o justificaci\'on estad\'istica alguna para establecer como condici\'on para identificar a un ganador que la distancia entre intervalos punteros fuese al menos de 0.6\%. En la p\'ag. 24 del citado informe del CTARCR simplemente se menciona respecto al \textit{Proceso de Estimaci\'on} lo siguiente:
\begin{quote}
  \noindent\textsl{``Para el proceso de estimaci\'on de las proporciones de votos por partido y coalici\'on, se especific\'o que se tendr\'ian como \textbf{reglas b\'asicas} las siguientes:}
	\begin{enumerate}
	  \item \textsl{Los m\'etodos de estimaci\'on utilizados deben llevar a conclusiones coherentes y comunes.}
		\item \textsl{Realizar estimaciones por intervalos de cuando menos una confiabilidad del 95\%.}
		\item \textsl{Para poder identificar un ganador, los intervalos de las primeras dos fuerzas contendientes deber\'an distar en al menos 0.6\%.''}
	\end{enumerate}
\end{quote}

\noindent Estas ``reglas b\'asicas'' que se adoptaron para la estimaci\'on del conteo r\'apido de dicha elecci\'on presentan diversos problemas. La Regla 1 es inespec\'ifica porque ?`c\'omo decidir si las conclusiones son ``coherentes'' y en qu\'e sentido? ?`A qu\'e se refiere con ``conclusiones comunes''? El informe no lo aclara. La Regla 2 no especifica el nivel de confiabilidad de las estimaciones, solo da un cota inferior para ello al decir ``cuando menos una confiabilidad del 95\%'', lo cual dej\'o la puerta abierta a la utilizaci\'on de distintos y arbitrarios niveles de confiabilidad por encima del 95\%, pudi\'endose llegar al absurdo de utilizar una confiabilidad del 100\% y con ello entregar como intervalo de estimaci\'on $[0\%,100\%].$ De hecho el informe tampoco aclara cu\'ales fueron los niveles de confiabilidad finalmente utilizados con cada uno de los tres m\'etodos de estimaci\'on, ni si fue el mismo en los tres casos. De la Regla 3 surge inmediatamente una pregunta: ?`A qu\'e se refiere con que los intervalos estimados para los dos punteros ``disten'' en al menos 0{.}6\%? Tampoco se incluye en dicho informe la f\'ormula que se utiliz\'o para calcular ``distancia entre intervalos'', mucho menos un fundamento estad\'istico para ello ni para fijar esa ``distancia'' m\'inima en 0{.}6\%, simplemente fue un ``acuerdo'' del CTARCR cuya fundamentaci\'on no fue documentada en el informe. Por supuesto que matem\'aticamente es posible definir alguna forma de calcular distancia entre intervalos (por ejemplo, utilizando la distancia de Hausdorff) pero para que esto pueda aplicarse en un procedimiento de inferencia estad\'istica requerir\'ia al menos una justificaci\'on t\'ecnica, que de haberla no fue documentada ni revelada.

\medskip

\begin{figure}[t]
\begin{center}
\includegraphics[width = 10cm, keepaspectratio]{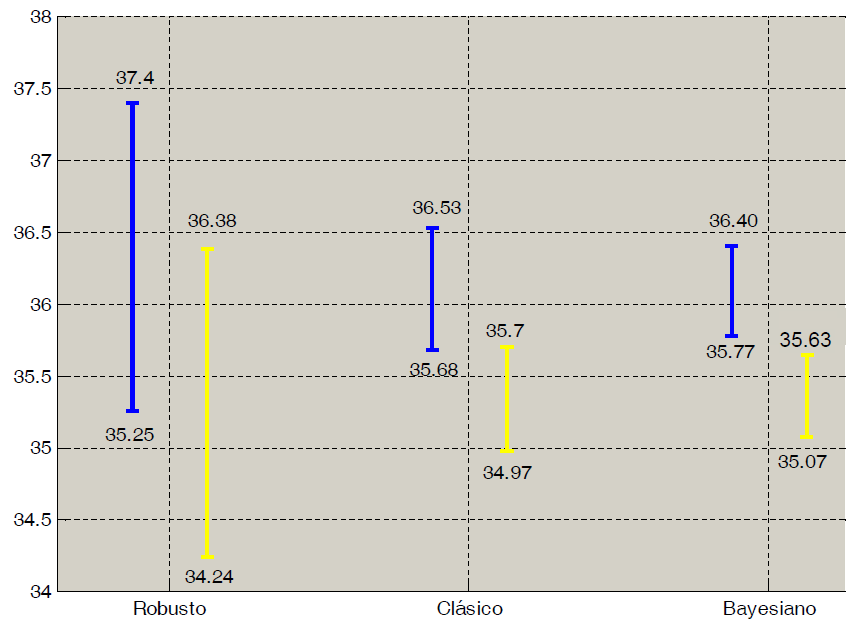}
\end{center}
\caption{Intervalos estimados para Felipe Calder\'on Hinojosa (FCH) en azul, y para Andr\'es Manuel L\'opez Obrador (AMLO) en amarillo bajo las metodolog\'ias denominadas Robusta, Cl\'asica y Bayesiana. Fuente: Informe del CTARCR al Instituto Federal Electoral (2006).}
\label{intervalos2006}
\end{figure}

\noindent Los intervalos estimados por el CTARCR la noche del 2 de julio de 2006 para los dos candidatos punteros (Felipe Calder\'on Hinojosa del PAN\footnote{Las siglas PAN corresponden a \textit{Partido Acci\'on Nacional}.} y Andr\'es Manuel L\'opez Obrador de la CPBT\footnote{Las siglas CPBT corresponden a \textit{Coalici\'on Por el Bien de Todos}.}), junto con el resultado del PREP\footnote{PREP son las siglas del \textit{Programa de Resultados Electorales Preliminares} que es un mecanismo de informaci\'on que, al t\'ermino de la jornada electoral, permite consultar, a trav\'es de internet, los resultados preliminares de las elecciones conforme los va recibiendo la autoridad electoral.} al cierre del mismo el 3 de julio\footnote{Fuente: http://prep2006.ife.org.mx/PREP2006/prep2006.html consultada el 17 de enero de 2018.} y el C\'omputo Distrital\footnote{Fuente: http://portalanterior.ine.mx/documentos/Estadisticas2006/presidente/nac.html consultada el 17 de enero de 2018.} del 5 de julio se presentan en el Cuadro \ref{resultadosCRPREPCD}, y una gr\'afica comparando dichos intervalos bajo los tres m\'etodos utilizados (Robusto, Cl\'asico y Bayesiano) se presenta en la Figura \ref{intervalos2006}.

\begin{table}[h]
\begin{center}
\begin{tabular}{|c|c|c|c|c|} \hline
  \textbf{M\'etodo}   &  \textbf{AMLO} & \textbf{Longitud intervalo} & \textbf{FCH} &  \textbf{Longitud intervalo}  \\ \hline
	Robusto             &  $[\,34{.}24\,,\,36{.}38\,]$  &         2.14       &  $[\,35{.}25\,,\,37{.}40\,]$  &  2.15  \\ \hline
	Cl\'asico           &  $[\,34{.}97\,,\,35{.}70\,]$  &         0.73       &  $[\,35{.}68\,,\,36{.}53\,]$  &  0.85  \\ \hline
	Bayesiano           &  $[\,35{.}07\,,\,35{.}63\,]$  &         0.56       &  $[\,35{.}77\,,\,36{.}40\,]$  &  0.63  \\ \hline
	PREP                &              35.34            &          --        &             36.38             &   --   \\ \hline
	C\'omputo Distrital &              35.31            &          --        &             35.89             &   --   \\ \hline
\end{tabular}
\end{center}
\caption{Intervalos estimados la noche del 2 de julio de 2006 por los m\'etodos estad\'isticos denominados por el CTARCR como Robusto, Cl\'asico y Bayesiano, el resultado al cierre del PREP el 3 de julio, y el resultado del C\'omputo Distrital el 5 de julio, para los dos candidatos punteros: Felipe Calder\'on Hinojosa (FCH) y Andr\'es Manuel L\'opez Obrador (AMLO).}
\label{resultadosCRPREPCD}
\end{table}

\medskip

\noindent La \textit{distancia de Hausdorff}, v\'ease por ejemplo Munkres (2002), para el caso de intervalos cerrados y acotados en el espacio m\'etrico de los n\'umeros reales ($\mathbb{R}$) bajo la m\'etrica usual en $\mathbb{R},$ resulta ser para un par de intervalos dados $\,I_1=[a_1,b_1]\,$ e $\,I_2=[a_2,b_2]\,$ como sigue:
\begin{equation}\label{distHausdorff}
  d_H(I_1,I_2)\,=\,\max\{\,|a_1-a_2|\,,\,|b_1-b_2|\,\}\,.
\end{equation}

\noindent En el Cuadro \ref{Hausdorff} se calcularon las distancias de Hausdorff (en puntos porcentuales) para los pares de intervalos estimados para los candidatos punteros bajo los tres m\'etodos utilizados para el conteo r\'apido de 2006 (ver Cuadro \ref{resultadosCRPREPCD}), y se agreg\'o, como simple referencia, la diferencia absoluta entre los puntos medios de dichos intervalos. Como se puede apreciar, bajo las tres metodolog\'ias estad\'isticas aplicadas, las distancias entre los pares de intervalos estimados para los dos candidatos punteros son en todos los casos superiores a 0.6\% y, por tanto, de acuerdo con la Regla 3 establecida por el propio comit\'e t\'ecnico asesor del conteo r\'apido (CTARCR), s\'i estuvieron en condiciones de identificar a un ganador de acuerdo con sus propias reglas autoimpuestas.

\begin{table}[h]
\begin{center}
\begin{tabular}{|c|c|c|} \hline
  \textbf{M\'etodo}   &  \textbf{Distancia de Hausdorff}  & \textbf{Diferencia entre puntos medios de intervalos} \\ \hline
	Robusto             & 1.010 & 1.015  \\ \hline
	Cl\'asico           & 0.710 & 0.770  \\ \hline
	Bayesiano           & 0.700 & 0.735  \\ \hline
\end{tabular}
\end{center}
\caption{Distancias de Hausdorff y entre puntos medios de los intervalos (en puntos porcentuales) estimados la noche del 2 de julio de 2006 por los m\'etodos estad\'isticos denominados por el CTARCR como Robusto, Cl\'asico y Bayesiano, para los dos candidatos punteros: Felipe Calder\'on Hinojosa (FCH) y Andr\'es Manuel L\'opez Obrador (AMLO).}
\label{Hausdorff}
\end{table}

\medskip

\noindent Respecto a la Regla 3, Aparicio (2009) interpreta la ``distancia'' de 0.6\% como \textit{margen de error} al mencionar:
\begin{quote}
  \noindent\textsl{``[\ldots] el margen de error del conteo r\'apido era de 0.6 por ciento pero como el resultado fue a\'un m\'as cerrado (0.58\%) no era posible declarar un ganador con los grados de confianza estad\'istica com\'unmente aceptados y que el mismo IFE hab\'ia anunciado previamente [\ldots] Si bien el conteo r\'apido levantado en 7636 casillas la noche del 2 de julio no apuntaba a un claro ganador (es decir, fuera de los m\'argenes de error del instrumento), \'este s\'i suger\'ia una elecci\'on con un margen menor a 0.6 por ciento de los votos, al tiempo que daba a Felipe Calder\'on una mayor probabilidad de aventajar en el resultado final. Con este dato era claro que el resultado observado al cierre del PREP la noche del 3 de julio, y que daba un margen de 1.04 por ciento a favor de Calder\'on, ten\'ia que reducirse al llegar al c\'omputo distrital --tal como ocurri\'o: el c\'omputo distrital arroj\'o un margen de s\'olo 0.58 por ciento, el cual valid\'o la estimaci\'on inicial del conteo r\'apido.''}
\end{quote}

\noindent Consultando el informe entregado por el CTARCR al IFE (2006) en la p\'agina 8 se estableci\'o lo siguiente:
\begin{quote}
  \noindent\textsl{``[\ldots] el objetivo fundamental es obtener un dise\~no y tama\~no de muestra que permitan estimar con un error aceptablemente peque\~no, del \textbf{0.5\%} \ldots''}
\end{quote}
\noindent y luego en la p\'agina 9:
\begin{quote}
  \noindent\textsl{``[\ldots] el Comit\'e evalu\'o el tama\~no de muestra de 7500 casillas a la luz de los errores de estimaci\'on que arrojar\'ia. Con ese tama\~no de muestra se tiene un error de alrededor de \textbf{0.3\%} \ldots''}
\end{quote}
\noindent De acuerdo con Eslava (2006), miembro de dicho CTARCR, con una muestra efectivamente recibida de 7236 casillas el error observado estimado fue de \textbf{0.42\%} para el caso del PAN (candidato Felipe Calder\'on Hinojosa) y de \textbf{0.36\%} para la CPBT (candidato Andr\'es Manuel L\'opez Obrador), por lo que tampoco resulta aceptable la interpretaci\'on del \textbf{0.6\%} de la Regla 3 como margen de error, tal cual se sugiere en Aparicio (2009).

\bigskip

\noindent Mendoza y Nieto-Barajas (2016) presentan un enfoque bayesiano para las estimaciones del conteo r\'apido de la elecci\'on presidencial de M\'exico de 2006 justamente, y es importante aclarar adem\'as que Mendoza tambi\'en form\'o parte del comit\'e t\'ecnico asesor para los conteos r\'apidos para dicha elecci\'on. El modelo bayesiano que publicaron, adem\'as de permitir las estimaciones por intervalo para las proporciones de votos que obtienen los candidatos, permite calcular la \textbf{probabilidad de triunfo}, y a las 10:15 p.m de la noche del d\'ia de la elecci\'on del 2 de julio de 2006 estimaron que la probabilidad de que el candidato del PAN (Felipe Calder\'on Hinojosa) fuese ganador era de \textbf{!`99.94\%!} C\'alculo que contrasta fuertemente con Eslava (2006) que concluy\'o que ``estad\'isticamente no era posible distinguir con un alto grado de confiabilidad a un candidato ganador''. Desafortunadamente la probabilidad de triunfo no se comunic\'o la noche de la elecci\'on, el comit\'e asesor del conteo r\'apido simplemente decidi\'o que no pod\'ia distinguir un ganador con base en un ``acuerdo previo'' sobre una determinada y arbitraria distancia interv\'alica que qui\'en sabe c\'omo se calcula ni qu\'e justificaci\'on probabil\'istica o estad\'istica tiene. En palabras de Mendoza y Nieto-Barajas:
\begin{quote}
  \noindent\textsl{``Our Bayesian model produced disjoint 99\% intervals, for the two leading parties, whose limits where really close to each other but separated. This fact is relevant because, non expert readers might think that the risk of a false winner call was large. This was not true and the evidence is provided precisely by the 22:15 dispersion diagram (bottom right panel in Fig. 4), where the probability of PAN being the winner is 0.9994. In fact, \textbf{the marginal intervals could intersect and nonetheless the relevant information to call a winner is provided by the corresponding joint bivariate posterior distribution}.''}
\end{quote}

\medskip

\noindent Lo anterior coincide con lo discutido en la secci\'on anterior respecto a calcular la probabilidad de triunfo y que el asunto de que los intervalos de estimaci\'on se intersecten o no deber\'ia quedar fuera de toda consideraci\'on porque no son probabil\'isticamente comparables, y porque en un momento dado la probabilidad de triunfo puede resultar elevada a pesar de existir traslape entre dichos intervalos. Como ya se mencion\'o, el informe que el CTARCR entreg\'o al IFE (2006) no especific\'o cu\'ales fueron los niveles de confiabilidad finalmente utilizados con cada uno de los tres m\'etodos de estimaci\'on, pero de Mendoza y Nieto-Barajas (2016) queda claro que al menos para el caso del m\'etodo Bayesiano fue de 99\%. Con esto \'ultimo fue posible realizar un an\'alisis similar al del Contraejemplo 3 de la secci\'on anterior (el c\'odigo de programaci\'on se anexa en el Ap\'endice, as\'i como un enlace para descarga del mismo, para fines de reproducibilidad) calculando bajo distintos niveles de dependencia tanto la probabilidad de triunfo de FCH (PAN) sobre AMLO (CPBT) como la probabilidad de que la diferencia entre ellos fuese menor a 0.6\%. Los resultados se resumen en el Cuadro \ref{analisis2006}, donde se aprecia que con probabilidades de triunfo muy elevadas en todos los casos, la probabilidad de que la diferencia fuese menor a 0.6\% pod\'ia llegar a niveles hasta de 27\% en un momento dado.

\medskip

\begin{table}[h]
\begin{center}
\begin{tabular}{|c|c|c|} \hline
  Correlaci\'on de Spearman & $\prob(\text{FCH}>\text{AMLO})$ en \% & $\prob(|\text{FCH}-\text{AMLO}| < 0{.}6\%)$ en \% \\ \hline
	-0.9 & 99.94  & 27.44 \\ \hline
	-0.6 & 99.98  & 25.76 \\ \hline
	-0.3 & 99.99+ & 23.52 \\ \hline
	 0.0 & 99.99+ & 20.43 \\ \hline
	+0.3 & 99.99+ & 15.97 \\ \hline
	+0.6 & 99.99+ &  9.18 \\ \hline
	+0.9 & 99.99+ &  0.42 \\ \hline
\end{tabular}
\end{center}
\caption{Probabilidad de triunfo de FCH sobre AMLO y probabilidad de que la diferencia entre ambos fuese menor a 0.6\% bajo distintos niveles de dependencia (correlaci\'on de Spearman y c\'opula Normal) y los intervalos de probabilidad 99\% estimados bajo el m\'etodo Bayesiano.}
\label{analisis2006}
\end{table}

\medskip

\begin{figure}[t]
\begin{center}
\includegraphics[width = 15cm, keepaspectratio]{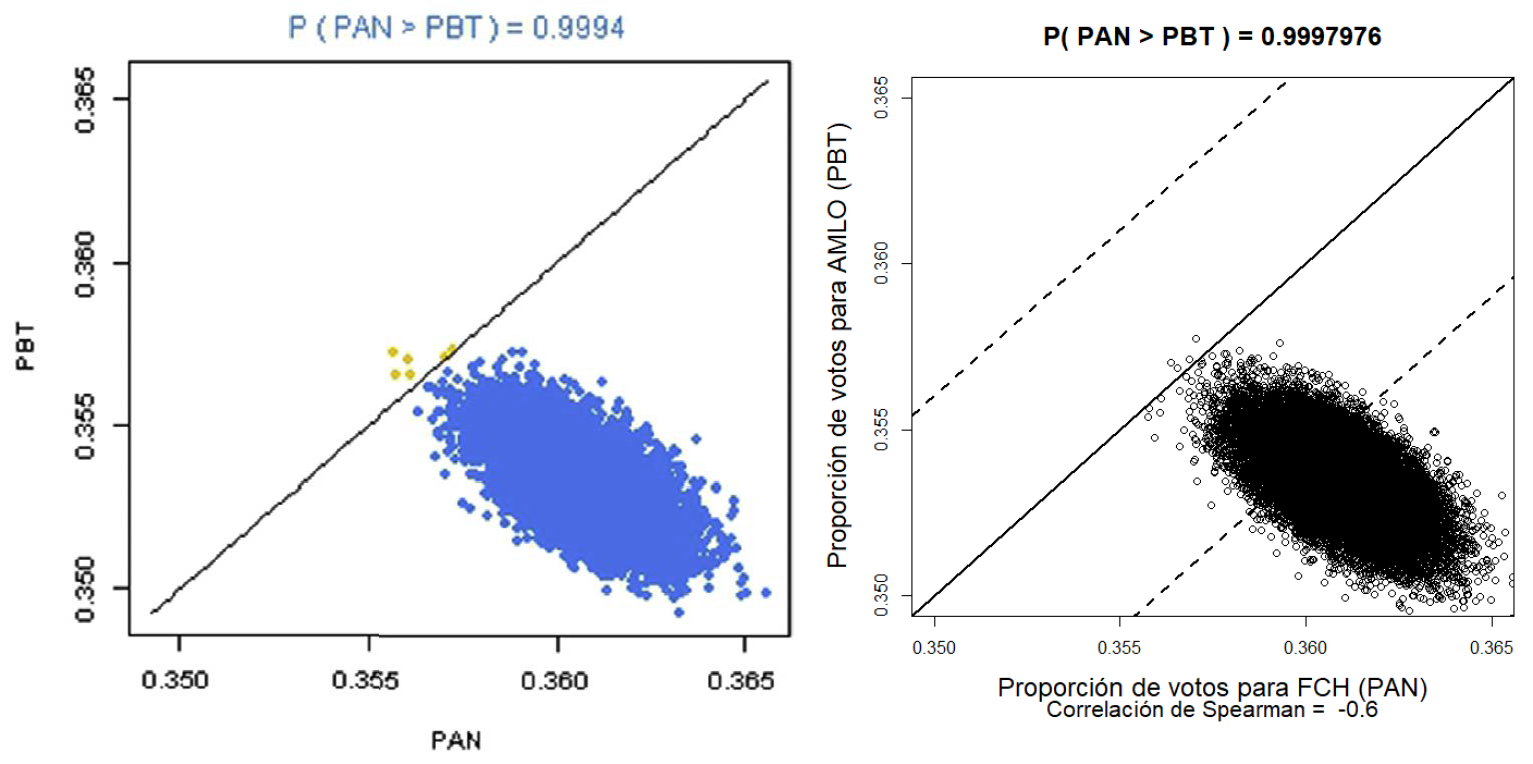}
\end{center}
\caption{Izquierda: Simulaciones conjuntas a posteriori FCH/AMLO (PAN/PBT) en Mendoza y Nieto-Barajas (2016). Derecha: Simulaciones conjuntas con c\'opula Normal, correlaci\'on de Spearman de -0.6 y utilizando los intervalos de probabilidad 99\% estimados por Mendoza y Nieto-Barajas (2016).}
\label{Mendoza}
\end{figure}

\noindent Mendoza y Nieto-Barajas (2016) no incluyeron en su trabajo el grado de dependencia entre los dos candidatos punteros, pero en la Figura \ref{Mendoza} se compara la gr\'afica de los autores donde simularon la distribuci\'on conjunta a posteriori de los candidatos punteros con la que estimaron la probabilidad de triunfo de FCH sobre AMLO (que a la vista implica una dependencia negativa) versus lo obtenido en el Cuadro \ref{analisis2006} con una correlaci\'on de Spearman igual a -0.6 con c\'opula Normal\footnote{Se utiliz\'o una c\'opula Normal (o Gaussiana) porque el m\'etodo Bayesiano utilizado en Mendoza y Nieto-Barajas (2016) se apoya en una distribuci\'on multivariada Normal -- Wishart invertida.} y los intervalos de probabilidad 99\% estimados bajo el m\'etodo Bayesiano. Esto nos da cierta idea de que a\'un con una probabilidad de triunfo de FCH sobre AMLO de casi 100\% la probabilidad de que la diferencia en porcentaje de votos entre ambos candidatos fuese menor a 0.6\% pod\'ia haber sido de aproximadamente 26\%. En las gr\'aficas de la Figura \ref{Mendoza}, la parte inferior a la l\'inea recta corresponde a escenarios donde FCH triunfa sobre AMLO, y la regi\'on encerrada entre las dos l\'ineas punteadas corresponde al caso en que la diferencia en puntos porcentuales entre ambos candidatos punteros es menor a 0.6\%.

\medskip

\medskip

\noindent Si en lo anterior se repitieran los c\'alculos pero utilizando los intervalos estimados por el m\'etodo Cl\'asico (ver Cuadro \ref{resultadosCRPREPCD}) manteniendo c\'opula Normal con correlaci\'on de Spearman de -0.6 se obtendr\'ian los resultados que se resumen en el Cuadro \ref{Clasico}. Como puede apreciarse, la probabilidad de triunfo de FCH sobre AMLO es similarmente elevada con probabilidades no despreciables de que la diferencia entre los candidatos punteros fuese menor a 0.6\%. Como el informe del comit\'e t\'ecnico asesor del conteo r\'apido no especific\'o el nivel de confianza utilizado para las estimaciones del m\'etodo Cl\'asico, se hizo el an\'alisis con diversos niveles de confianza. N\'otese que, a diferencia de los intervalos estimados por el m\'etodo Bayesiano, los intervalos del m\'etodo Cl\'asico s\'i tienen traslape (ver Cuadro \ref{resultadosCRPREPCD} y Figura \ref{intervalos2006}), y a pesar de ello la probabilidad de triunfo de FCH (PAN) es muy similar al caso Bayesiano, al igual que la probabilidad de que la diferencia entre candidatos fuese menor a 0.6\%.

\noindent

\begin{table}[h]
\begin{center}
\begin{tabular}{|c|c|c|} \hline
  Confianza del intervalo en \% & $\prob(\text{FCH}>\text{AMLO})$ en \% & $\prob(|\text{FCH}-\text{AMLO}| < 0{.}6\%)$ en \% \\ \hline
	 95.0 & 98.30  & 31.91 \\ \hline
	 99.0 & 99.74  & 26.82 \\ \hline
	 99.5 & 99.88  & 25.03 \\ \hline
	 99.9 & 99.99+ & 17.53 \\ \hline
\end{tabular}
\end{center}
\caption{Probabilidad de triunfo de FCH sobre AMLO y probabilidad de que la diferencia entre ambos fuese menor a 0.6\% con correlaci\'on de Spearman de -0.6, c\'opula Normal y los intervalos de probabilidad estimados bajo el m\'etodo Cl\'asico, suponiendo distintos niveles de confianza de dichos intervalos.}
\label{Clasico}
\end{table}

\bigskip

\begin{figure}[h]
\begin{center}
\includegraphics[width = 15cm, keepaspectratio]{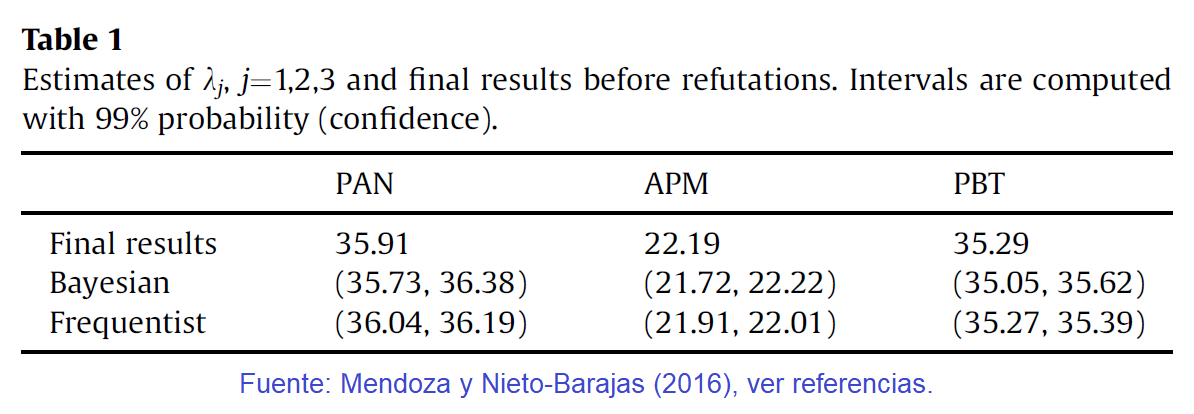}
\end{center}
\caption{\textit{Table 1} en Mendoza y Nieto-Barajas (2016) con intervalos al mismo nivel de probabilidad (confianza) de 99\% para los m\'etodos Bayesiano y Cl\'asico (o Frecuentista).}
\label{Tabla1Mendoza}
\end{figure}

\noindent En congruencia con el nivel de confianza utilizado para decidir el tama\~no de muestra, el nivel de confianza de los intervalos estimados debiera ser tambi\'en del 95\%. En el caso de la elecci\'on presidencial de M\'exico en el a\~no 2006, el informe final del comit\'e t\'ecnico asesor para el conteo r\'apido (IFE, 2006) no especific\'o el nivel de confianza utilizado para los intervalos calculados bajo los tres m\'etodos utilizados (denominados Robusto, Cl\'asico y Bayesiano), pero de acuerdo a lo publicado al respecto por Mendoza y Nieto-Barajas (2016), al menos para el m\'etodo Bayesiano fueron intervalos de probabilidad 99\%, ver Figura \ref{Tabla1Mendoza} (que corresponde a \textit{Table 1} en Mendoza y Nieto-Barajas, 2016).

\medskip

\medskip

\begin{figure}[h]
\begin{center}
\includegraphics[width = 11cm, keepaspectratio]{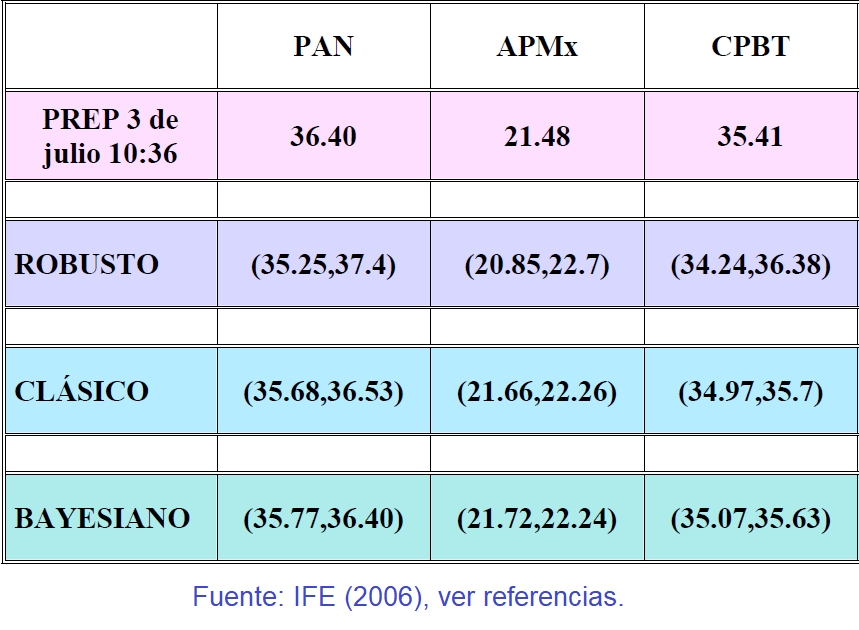}
\end{center}
\caption{Resultados del conteo r\'apido de la elecci\'on presidencial de M\'exico en 2006, de acuerdo al informe del CTARCR al IFE (2006).}
\label{CR2006B}
\end{figure}

\noindent Para comparar su m\'etodo Bayesiano, Mendoza y Nieto-Barajas (2016) presentan los intervalos de confianza 99\% que corresponder\'ian al m\'etodo Cl\'asico (o Frecuentista). Lo que es de notarse es que las longitudes de los intervalos del m\'etodo Cl\'asico (o Frecuentista) resultaron de 0.15 (PAN), 0.10 (APM) y 0.12 (PBT), longitudes mucho menores que las de los intervalos reportados por el CTARCR al IFE (2006) para el m\'etodo Cl\'asico (ver Figura \ref{CR2006B}) que fueron 0.85 (PAN), 0.60 (APM) y 0.73 (PBT), lo cual deja al descubierto que el nivel de confianza de los intervalos calculados con el m\'etodo Cl\'asico en el conteo r\'apido de 2006 fue a\'un superior al 99\%, quiz\'as algo as\'i como 99.9\%.

\medskip

\medskip

\begin{figure}[h]
\begin{center}
\includegraphics[width = 11cm, keepaspectratio]{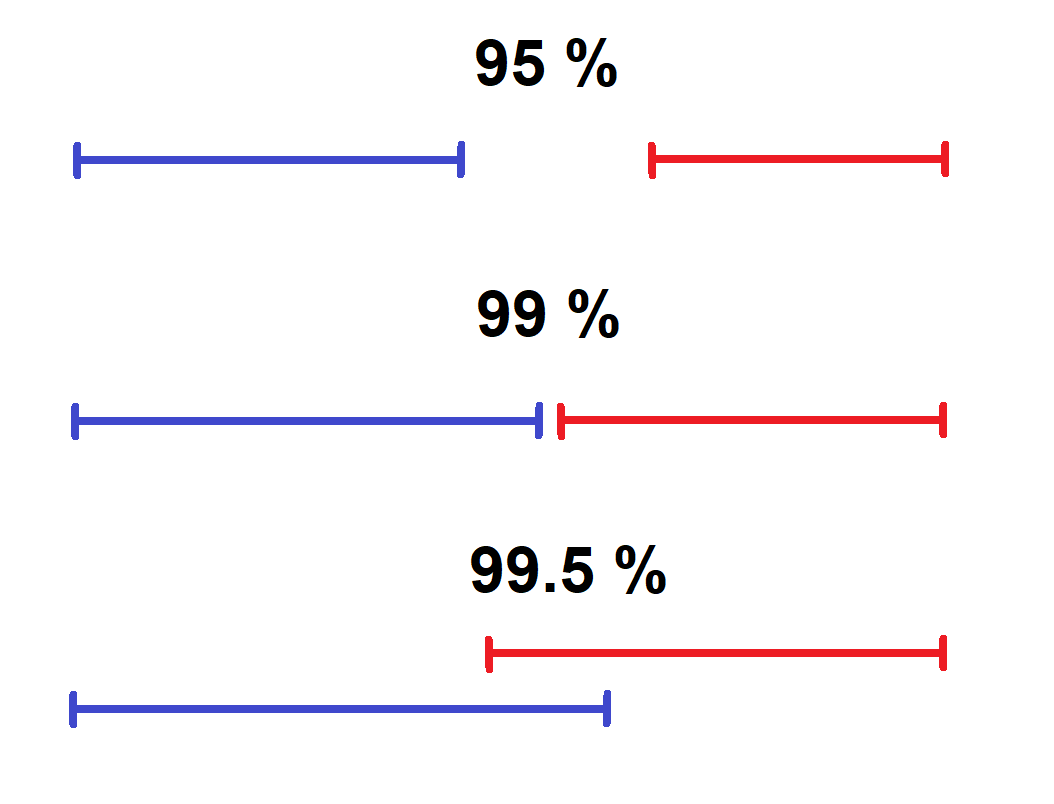}
\end{center}
\caption{Con la misma informaci\'on muestral es posible lograr que intervalos estimados se separen o traslapen modificando el nivel de confianza para calcularlos. \textit{Fuente:} Elaboraci\'on propia.}
\label{intervalos10}
\end{figure}

\noindent La longitud de un intervalo de confianza es funci\'on creciente del nivel de confianza elegido, esto es, a mayor/menor nivel de confianza se obtiene una mayor/menor longitud del intervalo. En el informe del CTARCR al IFE (2006) qued\'o asentado que dicho comit\'e t\'ecnico acord\'o, entre otras cosas, que el nivel de confianza de los intervalos ser\'ia de al menos 95\%. Pero al decir ``al menos'' dej\'o totalmente abierta la puerta a la discrecionalidad de quienes realizaron los c\'alculos de los intervalos, que ni siquiera utilizaron el mismo nivel de confiabilidad (99\% para el m\'etodo Bayesiano, 99.9\% aproximadamente para el m\'etodo Cl\'asico). Y es que precisamente con discrecionalidad sobre el nivel de confianza a utilizar se puede lograr que los intervalos se separen o traslapen a placer, como se ilustra en la Figura \ref{intervalos10}.

\medskip

\bigskip

\noindent Los intervalos estimados por el m\'etodo denominado ``Robusto'' no son dignos de an\'alisis alguno por la sencilla raz\'on de que dicho m\'etodo supone que la muestra ha sido seleccionada de acuerdo con un esquema aleatorio simple, lo cual resulta incongruente con el dise\~no de muestreo utilizado para dicho conteo r\'apido que fue \textit{estratificado}. La idea de aplicar un \textit{muestreo aleatorio estratificado} radica en el beneficio inferencial que se obtiene por establecer o escoger una estratificaci\'on tal que en cada estrato sea razonable suponer cierta homogeneidad en estas subpoblaciones, en contraste con la heterogeneidad que se tendr\'ia al obtener muestras directamente de la poblaci\'on total sin estratificar, en deterioro de la precisi\'on de las inferencias que se desea realizar, v\'ease por ejemplo S\"{a}rndal \textit{et al.} (1992). Es por ello que la longitud de los intervalos estimados por el m\'etodo Robusto es considerablemente mayor que la de los m\'etodos Bayesiano y Cl\'asico, v\'ease Figura \ref{intervalos2006}. En el citado informe del CTARCR, respecto al m\'etodo Robusto se afirma en la p\'ag. 114:
\begin{quote}
  \noindent\textsl{``[\ldots] si se logra tener una gran parte de la muestra prevista, los intervalos producidos deber\'an ser muy parecidos a los obtenidos con los m\'etodos Bayesiano y cl\'asico.''}
\end{quote}

\noindent Como era de esperarse, no ocurri\'o as\'i. A pesar de que se logr\'o recabar 95.12\% del total de la muestra prevista para el conteo r\'apido, 99.8\% de los estratos y 100\% de los distritos (p\'ag. 22 de dicho informe) es evidente que los intervalos producidos por el m\'etodo ``Robusto'' distan mucho de los obtenidos por los m\'etodos Bayesiano y Cl\'asico. En el Cuadro \ref{matrizH} se presenta una matriz de distancias de Hausdorff (\ref{distHausdorff}) entre los intervalos generados por los tres m\'etodos para los dos candidatos punteros, correspondiendo la matriz triangular superior a los intervalos estimados para FCH (PAN) y la inferior lo correspondiente a AMLO (PBT). Mientras que las distancias de Hausdorff entre intervalos de los m\'etodos Cl\'asico y Bayesiano están en el rango 0.09--0.10\%, la distancia del m\'etodo Robusto respecto a los otros dos m\'etodos oscila en 0.43--0.83\%.

\begin{table}[h]
\begin{center}
\begin{tabular}{|c|c|c|c|} \hline
  {distancia Hausdorff}   & Robusto & Cl\'asico & Bayesiano \\ \hline
	Robusto             &    --   &    0.43   &    0.52   \\ \hline
	Cl\'asico           &  0.73   &     --    &    0.09   \\ \hline
	Bayesiano           &  0.83   &    0.10   &     --    \\ \hline
\end{tabular}
\end{center}
\caption{Distancias de Hausdorff entre intervalos estimados bajo los m\'etodos Robusto, Cl\'asico y Bayesiano. Los valores de la matriz triangular superior corresponden a los intervalos estimados para FCH (PAN), y la inferior lo correspondiente a AMLO (PBT).}
\label{matrizH}
\end{table}

\section{Conclusiones}

\noindent Se demostr\'o, tanto de forma te\'orica como mediante el an\'alisis de un conteo r\'apido, que la \textbf{probabilidad de triunfo} electoral de un candidato puntero puede resultar muy cercana a 100\% a pesar de que exista traslape de los intervalos estimados para la proporci\'on de votos de los dos candidatos punteros. Declarar ``empate t\'ecnico'' porque hay traslape de intervalos no tiene fundamento probabil\'istico ni justificaci\'on estad\'istica, y por tanto no es \'util para interpretar encuestas o conteos r\'apidos electorales, ya que dichos intervalos se calculan de forma individual sin tomar en cuenta la dependencia probabil\'istica entre candidatos, que resulta fundamental para contestar preguntas que involucran simult\'aneamente a dos o m\'as de ellos, como por ejemplo que alguno triunfe sobre otro.

\medskip

\noindent En el caso particular de M\'exico, el \textit{Reglamento de Elecciones} del INE (2016) en su art\'iculo 380, inciso 3, establece lo siguiente:
\begin{quote}
  \noindent\textsl{``Sea cual fuere la muestra recabada y los resultados obtenidos, el COTECORA\footnote{COTECORA significa Comit\'e T\'ecnico para los Conteos R\'apidos.} deber\'a presentar un reporte al Consejo General u \'Organo Superior de Direcci\'on que corresponda, en el que indique, adem\'as, las condiciones bajo las cuales se obtuvieron los resultados, as\'i como \textbf{las conclusiones que de ellos puedan derivarse}. Las estimaciones deber\'an presentarse en forma de intervalos de confianza para cada contendiente.''}
\end{quote}
\noindent Esta redacci\'on no establece que las conclusiones deban derivarse exclusivamente de los intervalos de confianza, sino de los resultados obtenidos --en un sentido m\'as general-- a partir de la muestra recabada, y dentro de ello cabe incluir el c\'alculo de la \textbf{probabilidad de triunfo} del candidato puntero con base en dicha informaci\'on muestral, y no necesariamente a partir del criterio de que exista o no traslape de dichos intervalos, que se demostr\'o no tiene sustento estad\'istico. Si, por ejemplo, se reportara que determinado candidato result\'o con probabilidad 99\% de ser triunfador de la elecci\'on, se entiende (y puede explicarse as\'i) que existe una probabilidad de 1\% de que no resulte finalmente as\'i en el recuento total de votos (c\'omputo distrital). 

\medskip

\noindent La \textbf{probabilidad de triunfo} es, como se explic\'o en la introducci\'on del presente art\'iculo citando a Lindley (2000), una medida del grado de certidumbre/incertidumbre sobre un evento de inter\'es (el triunfo de un candidato en este caso) con base en la informaci\'on disponible en un momento dado, y es as\'i como podr\'ia explicarse a no expertos, porque es justamente lo que puede ofrecer la estad\'istica con base en informaci\'on parcial (muestral), ya que la certidumbre total solo es posible con acceso a la totalidad de la informaci\'on.

\section*{Referencias}



\noindent Aparicio, J. (2009) An\'alisis estad\'istico de la elecci\'on presidencial de 2006 ?`fraude o errores aleatorios? \textit{Pol\'itica y Gobierno, Volumen Tem\'atico Elecciones en M\'exico} \textbf{2}, 225--243.\medskip

\noindent Campos, R., Penna, C. (2004) ?```Empate t\'ecnico'' o ``too close to call''? www.consulta.mx [consultado el 13-Ene-2018]\medskip

\noindent Embrechts, P., McNeil, A.J., Strauman, D. (1999) Correlation: pitfalls and alternatives. \textit{Risk Magazine} \textbf{5}, 69--71.\medskip


\noindent Eslava, G. (2006) Las elecciones de 2006, un an\'alisis del conteo r\'apido. \textit{Ciencias} \textbf{84}, octubre-diciembre, 30--37. [En l\'inea]\medskip

\noindent Frank, M.J. (1979) On the simultaneous associativity of $F(x,y)$ and $x+y-F(x,y).$ \textit{Aequationes Math.} \textbf{19}, 194--226.\medskip

\noindent Fr\'echet, M. (1951) Sur les tableaux de corr\'elation dont les marges sont donn\'ees. \textit{Ann. Univ. Lyon} \textbf{14}, (Sect. A Ser. 3), 53--77.\medskip

\noindent Hoeffding, W. (1940) Masstabinvariante Korrelationstheorie. \textit{Schriften des Matematischen Instituts und des Instituts f\"ur Angewandte Mathematik der Universit\"at Berlin} \textbf{5}, 179--223.\medskip

\noindent Hofert, M., Kojadinovic, I., Maechler, M., Yan, J. (2017) \texttt{copula}: \textit{Multivariate Dependence with Copulas.} R
  package version 0.999-18 URL https://CRAN.R-project.org/package=copula\medskip
	
\noindent Instituto Federal Electoral IFE (2006) \textit{Informe sobre las actividades del Comit\'e T\'ecnico Asesor para la Realizaci\'on de Conteos R\'apidos}, 147 pp. www.portalanterior.ine.mx [consultado el 17 de diciembre de 2017]\medskip

\noindent Instituto Nacional Electoral INE (2016) \textit{Reglamento de Elecciones} (M\'exico). www.ine.mx [consultado el 12 de enero de 2018]\medskip

\noindent Lindley, D.V. (2000) The Philosophy of Statistics. \textit{Journal of the Royal Statistical Society. Series D (The Statistician)} \textbf{49} (3), 293--337.\medskip

\noindent Mendoza, M., Nieto-Barajas, L.E. (2016) Quick counts in the Mexican presidential elections: A Bayesian approach. \textit{Electoral Studies} \textbf{43}, 124--132.\medskip


\noindent Munkres, J.R. (2002) \textit{Topolog\'ia.} Prentice Hall.\medskip

\noindent Nelsen, R.B. (2006) \textit{An Introduction to Copulas.} Springer.\medskip


\noindent R Core Team (2017) R: A language and environment for statistical computing. \textit{R Foundation for Statistical Computing},
  Vienna, Austria. URL https://www.R-project.org/\medskip
	
\noindent S\"{a}rndal, C.-E., Swensson, B., Wretman, J. (1992) \textit{Model Assisted Survey Sampling.} Springer.\medskip

\noindent Sklar, A. (1959) Fonctions de r\'epartition \`a $n$ dimensions et leurs marges. \textit{Publ. Inst. Statist. Univ. Paris,} \textbf{8}, 229--231.\medskip


\section*{Ap\'endice}

\noindent C\'odigo en lenguaje de programaci\'on \texttt{R} para los c\'alculos del \textbf{Contraejemplo 3}, tambi\'en descargable de\linebreak https://goo.gl/z5FYNg

\medskip

\begin{verbatim}
# Construccio'n de intervalo para X
gama <- 0.05 # la probabilidad del intervalo sera' 1-gama
x.sup <- 0.38 # extremo superior del intervalo
epsilon.X <- 0.03 # margen de error 
x.inf <- x.sup - 2*epsilon.X # extremo inferior del intervalo
# Estimacio'n de para'metros alfa y beta: 
hX <- function(ab) (qbeta(1 - gama/2, ab[1], ab[2]) - 2*x.sup)^2 +
                   (qbeta(gama/2, ab[1], ab[2]) - 2*x.inf)^2
(hX.optim <- nlm(hX, c(8, 1)))
# Comprobando probabilidad del intervalo:
pbeta(2*x.sup, hX.optim$estimate[1], hX.optim$estimate[2]) -
pbeta(2*x.inf, hX.optim$estimate[1], hX.optim$estimate[2])
# Graficar densidad de X y su intervalo:
x <- seq(0, 1/2, length.out = 1000)
plot(x, 2*dbeta(2*x, hX.optim$estimate[1], hX.optim$estimate[2]),
     type = "l", xlab = "proporcio'n de votos", ylab = "densidad de probabilidades",
     xlim = c(0.2, 0.45), col = "blue", lwd = 3, cex.lab = 1.5, cex.sub = 1.3)
rug(c(x.inf, x.sup), lwd = 3, col = "blue")
cat("Intervalo para X: [", x.inf, ",", x.sup, "]", "\n")
# Construccio'n de intervalo para Y
epsilon.Y <- 0.03 # margen de error
tau <- 0.005 # traslape
y.sup <- x.inf + tau # extremo superior del intervalo
y.inf <- y.sup - 2*epsilon.Y # extremo inferior del intervalo
# Estimacio'n de para'metros alfa y beta:
hY <- function(ab) (qbeta(1 - gama/2, ab[1], ab[2]) - 2*y.sup)^2 +
                   (qbeta(gama/2, ab[1], ab[2]) - 2*y.inf)^2
(hY.optim <- nlm(hY, c(8, 1)))
# Comprobando probabilidad del intervalo:
pbeta(2*y.sup, hY.optim$estimate[1], hY.optim$estimate[2]) -
pbeta(2*y.inf, hY.optim$estimate[1], hY.optim$estimate[2])
# Graficar densidad de Y y su intervalo:
lines(x, 2*dbeta(2*x, hY.optim$estimate[1], hY.optim$estimate[2]), col = "red", lwd = 3)
rug(c(y.inf, y.sup), lwd = 3, col = "red")
title("Intervalos traslapados", cex.main = 1.5)
cat("Intervalo para Y: [", y.inf, ",", y.sup, "]", "\n")
# Dependencia para (X,Y): Co'pula Frank
library(copula) # Paquete <copula> previamente instalado
rho <- -0.5 # valor de correlacio'n de Spearman (en [-1,+1])
(th <- iRho(frankCopula(), rho)) # calculando para'metro
Copula.Frank <- frankCopula(th) # definiendo la co'pula
n <- 1000000 # tamanio de la simulacio'n
set.seed(666) # especificando semilla para reproducibilidad
uv.sim <- rCopula(n, Copula.Frank) # simulando pseudo-observaciones
# Agregando dependencia a (X,Y):
xy.sim <- matrix(nrow = n, ncol = 2)
xy.sim[ , 1] <- (1/2)*qbeta(uv.sim[ , 1], hX.optim$estimate[1], hX.optim$estimate[2])
xy.sim[ , 2] <- (1/2)*qbeta(uv.sim[ , 2], hY.optim$estimate[1], hY.optim$estimate[2])
cor(xy.sim, method = "spearman")[1, 2] # comprobando correlacio'n obtenida
# Comprobando estimaciones puntuales y por intervalo:
quantile(xy.sim[ , 1], probs = c(gama/2, 1/2, 1-gama/2))
quantile(xy.sim[ , 2], probs = c(gama/2, 1/2, 1-gama/2))
# Estimando probabilidad de triunfo de X sobre Y:
(P.triunfoX <- mean(xy.sim[ , 1] > xy.sim[ , 2]))
# Graficar parte de las simulaciones de (X,Y)
dev.new()
plot(xy.sim[1:3000, ], xlab = "Proporción de votos para  X", 
     ylab = "Proporcio'n de votos para  Y", main = "Candidatos punteros",
     sub = paste("Correlacio'n de Spearman = ", rho), cex.lab = 1.5,
     cex.main = 1.5, cex.sub = 1.3, xlim = c(.3,.4), ylim = c(.23,.36))
text(0.325, 0.235, paste("P(X > Y) =", P.triunfoX), cex = 2)
abline(a = 0, b = 1, lwd = 2)
\end{verbatim}

\bigskip

\noindent C\'odigo en lenguaje de programaci\'on \texttt{R} para los c\'alculos sobre la \textbf{elecci\'on presidencial de 2006}, tambi\'en descargable de https://goo.gl/TKoctG

\begin{verbatim}
# Construccio'n de intervalo para FCH (PAN)
gama <- 0.01 # la probabilidad del intervalo sera' 1-gama
x.inf <- 0.3577 # extremo inferior del intervalo
x.sup <- 0.3640 # extremo superior del intervalo
epsilon.X <- (x.sup - x.inf)/2 # margen de error 
# Estimacio'n de para'metros alfa y beta: 
hX <- function(ab) (qbeta(1 - gama/2, ab[1], ab[2]) - 2*x.sup)^2 +
                   (qbeta(gama/2, ab[1], ab[2]) - 2*x.inf)^2
(hX.optim <- nlm(hX, c(18, 18)))
# Comprobando probabilidad del intervalo:
pbeta(2*x.sup, hX.optim$estimate[1], hX.optim$estimate[2]) -
pbeta(2*x.inf, hX.optim$estimate[1], hX.optim$estimate[2])
# Graficar densidad de FCH y su intervalo:
x <- seq(0, 1/2, length.out = 10000)
plot(x, 2*dbeta(2*x, hX.optim$estimate[1], hX.optim$estimate[2]),
     type = "l", xlab = "proporcio'n de votos", ylab = "densidad de probabilidades",
     xlim = c(0.345, 0.365), col = "blue", lwd = 3, cex.lab = 1.5, cex.sub = 1.3,
     ylim = c(0, 400))
rug(c(x.inf, x.sup), lwd = 3, col = "blue")
cat("Intervalo FCH: [", x.inf, ",", x.sup, "]", "\n")
# Construccio'n de intervalo para AMLO (PBT)
y.inf <- 0.3507 # extremo inferior del intervalo
y.sup <- 0.3563 # extremo superior del intervalo
epsilon.Y <- (x.sup - x.inf)/2 # margen de error
# Estimacio'n de para'metros alfa y beta:
hY <- function(ab) (qbeta(1 - gama/2, ab[1], ab[2]) - 2*y.sup)^2 +
                   (qbeta(gama/2, ab[1], ab[2]) - 2*y.inf)^2
(hY.optim <- nlm(hY, c(18, 18)))
# Comprobando probabilidad del intervalo:
pbeta(2*y.sup, hY.optim$estimate[1], hY.optim$estimate[2]) -
pbeta(2*y.inf, hY.optim$estimate[1], hY.optim$estimate[2])
# Graficar densidad de PBT y su intervalo:
lines(x, 2*dbeta(2*x, hY.optim$estimate[1], hY.optim$estimate[2]), col = "red", lwd = 3)
rug(c(y.inf, y.sup), lwd = 3, col = "red")
title("AMLO (rojo) versus FCH (azul)", cex.main = 1.5)
cat("Intervalo para AMLO: [", y.inf, ",", y.sup, "]", "\n")
# Dependencia para (FCH,AMLO): Co'pula Normal
library(copula) # Paquete <copula> previamente instalado
rho <- -0.6 # valor de correlacio'n de Spearman (en [-1,+1])
(th <- iRho(normalCopula(), rho)) # calculando para'metro
Copula.Normal <- normalCopula(th) # definiendo la co'pula
n <- 10000000 # tamanio de la simulacio'n
set.seed(777) # especificando semilla para reproducibilidad
uv.sim <- rCopula(n, Copula.Normal) # simulando pseudo-observaciones
# Agregando dependencia a (X,Y):
xy.sim <- matrix(nrow = n, ncol = 2)
xy.sim[ , 1] <- (1/2)*qbeta(uv.sim[ , 1], hX.optim$estimate[1], hX.optim$estimate[2])
xy.sim[ , 2] <- (1/2)*qbeta(uv.sim[ , 2], hY.optim$estimate[1], hY.optim$estimate[2])
cor(xy.sim, method = "spearman")[1, 2] # comprobando correlacio'n obtenida
# Comprobando estimaciones puntuales y por intervalo:
quantile(xy.sim[ , 1], probs = c(gama/2, 1/2, 1-gama/2))
quantile(xy.sim[ , 2], probs = c(gama/2, 1/2, 1-gama/2))
# Estimando probabilidad de triunfo de X sobre Y:
(P.triunfoX <- mean(xy.sim[ , 1] > xy.sim[ , 2]))
# Estimando probabilidad de diferencia menor a 0.6%
(P.diferencia <- mean(abs(xy.sim[ , 1] - xy.sim[ , 2]) < 0.006))
# Graficar parte de las simulaciones de (X,Y)
dev.new()
plot(xy.sim[1:30000, ], xlab = "Proporcio'n de votos para FCH (PAN)", 
     ylab = "Proporcio'n de votos para AMLO (PBT)", main = paste("P( PAN > PBT ) =", P.triunfoX),
     sub = paste("Correlacio'n de Spearman = ", rho), cex.lab = 1.5,
     cex.main = 1.5, cex.sub = 1.3, xlim = c(.35,.365), ylim = c(.35,.365))
abline(a = 0, b = 1, lwd = 2)
abline(a = -0.006, b = 1, lwd = 2, lty = "dashed")
abline(a = 0.006, b = 1, lwd = 2, lty = "dashed")
\end{verbatim}

\end{document}